\documentclass[twocolumn, appendixfloats, numberedappendix, iop]{openjournal}

\usepackage[pass]{geometry}
\usepackage{natbib}
\usepackage{longtable}
\usepackage{graphicx}
\usepackage{threeparttable}
\usepackage[colorlinks=true,
            linkcolor=red,
            citecolor=blue,
            urlcolor=blue]{hyperref}
\let\oldhref\href

\renewcommand{\href}[2]{\oldhref{#1}{\textbf{#2}}}
\usepackage{xcolor}
\usepackage{orcidlink}

\newcommand\kms{km~s$^{-1}$}
\newcommand\nsample{1,901,834}
\newcommand\numNeV{35,978}
\newcommand{\OR}{log ([O III] 5008 / [O II] 3727 + 3729)}
\newcommand{\ORs}{log\,([O\,III]/[O\,II])}

\newcommand{\ORcorrs}{log ([O III]$_{\mathrm{corr}}$ / [O II]$_{\mathrm{corr}}$)}
\newcommand{\OIII}{[O\,III]\,5008}
\newcommand{\OII}{[O\,II] $\lambda\lambda$3727,3729}
\newcommand{\OIIp}{[O\,II] $\lambda\lambda$3727+3729}
\newcommand{\NII}{[N II] 6585}
\newcommand{\Ha}{H$\alpha$}
\newcommand{\Hb}{H$\beta$}
\newcommand{\NeV}{[Ne~V]~$\lambda\lambda$3347,3427}
\newcommand{\NeVp}{[Ne~V] 3347+3427}
\newcommand{\NeIII}{[Ne III] 3869}
\newcommand{\Ntwo}{log ([N II] 6585 / H$\alpha$)}
\newcommand{\NeR}{log ([Ne V] 3427 / [Ne III] 3869)}
\newcommand{\NeRs}{[Ne V]/[Ne III]} 
\newcommand{\Rthree}{log ([O III] 5008 / H$\beta$)}
\newcommand{\drake}{D. Miller et al. in prep.}
\newcommand{\numNeVgal}{33,817}
\newcommand{\OIIcorr}{[O II] 3727 + 3729$_{\mathrm{corr}}$}
\newcommand{\OIIIcorr}{([O III] 5008)$_{\mathrm{corr}}$}
\newcommand{\NeIIIcorr}{([Ne III] 3869)$_{\mathrm{corr}}$}

\newcommand{\NeRcorrs}{log ([Ne V] / [Ne III]$_{\mathrm{corr}}$)}
\newcommand{\Ledd}{$\log\,\lambda_{\mathrm{Edd}}$}
\newcommand{\edd}{$\lambda_{\mathrm{Edd}}$}
\newcommand\nsampledap{1,893,695}

\definecolor{note}{RGB}{200,0,0}

\shorttitle{The \NeV~AGN Diagnostic}
\shortauthors{O. Matthews Acu\~{n}a et al.}

\begin{document}

\title{\vspace{-0.25cm}The [Ne V] AGN Diagnostic in SDSS-IV/eBOSS: An Anticorrelation Between Ionization State and AGN Luminosity in Low-Redshift Coronal-Line Galaxies\vspace{-1.5cm}}

\author{Owen S. Matthews Acu\~{n}a*\orcidlink{0000-0001-9225-972X}$^{1}$,
Christy A. Tremonti\orcidlink{0000-0003-3097-5178}$^{1}$,
Nikko J. Cleri\orcidlink{0000-0001-7151-009X}$^{2,3,4}$,
Kyle B. Westfall\orcidlink{0000-0003-1809-6920}$^{5}$,
Bee R. Erena\orcidlink{0009-0005-3574-0548}$^{1}$, 
Jacob B. Stimac\orcidlink{0009-0006-0477-526X}$^{6}$,
Britt Lundgren\orcidlink{0000-0002-6463-2483}$^{6}$,
Drake Miller III\orcidlink{0009-0006-2178-1178}$^{7}$,
Aleksandar M. Diamond-Stanic$^{8}$\vspace{0.5em}
}
\affiliation{
  $^{1}$Department of Astronomy, University of Wisconsin-–Madison, Madison, WI 53706, USA\\
  $^{2}$Department of Astronomy and Astrophysics, The Pennsylvania State University, University Park, PA 16802, USA\\
  $^{3}$Institute for Computational and Data Sciences, The Pennsylvania State University, University Park, PA 16802, USA\\
  $^{4}$Institute for Gravitation and the Cosmos, The Pennsylvania State University, University Park, PA 16802, USA\\
  $^{5}$University of California Observatories, University of California, Santa Cruz, 1156 High Street, Santa Cruz, CA 95064, USA\\
  $^{6}$Department of Physics and Astronomy, University of North Carolina Asheville, Asheville, NC 28804, USA\\
  $^{7}$Department of Astrophysical and Planetary Sciences, University of Colorado Boulder, Boulder, CO 80309, USA\\
  $^{8}$Department of Physics and Astronomy, Bates College, Lewiston, ME 04240, USA\\
}
\thanks{Corresponding author: Owen S. Matthews Acu\~{n}a:\\ \href{matthewsacun@wisc.edu}{matthewsacun@wisc.edu}}

\begin{abstract}
Traditional narrow-line galaxy classification diagnostics fail at high redshift, where low metallicity drives star-forming galaxies and Active Galactic Nuclei (AGN) into overlapping regions of the Baldwin--Phillips--Terlevich (BPT) diagram.
Coronal lines, with ionization potentials exceeding 100~eV, offer a robust alternative: they cannot be produced by normal stellar populations, arising almost exclusively from AGN or fast radiative shocks, and are insensitive to abundance evolution.
Among these, the \NeV~doublet is the brightest optical coronal line tracer of AGN activity.
Existing coronal line catalogs contain only $\sim10^{3}$ objects, too few for a statistical assessment of the completeness and purity of [Ne~V]-selected AGN samples.  We identify \numNeVgal~galaxies with \NeV~S/N $>$ 5 at $z =  0.147 - 1.12$ in the Sloan Digital Sky Survey IV extended Baryon Oscillations Spectroscopic Survey (SDSS-IV/eBOSS), exceeding the combined literature total by more than an order of magnitude.
[Ne~V]~preferentially selects rapidly accreting, relatively unobscured AGN, with only 41.5\% of eBOSS BPT AGN showing detectable [Ne~V] emission.
By stacking spectra that lack [Ne~V] on a grid of black hole mass and \OIII~luminosity, we recover [Ne~V] in the majority of bins. This implies that the non-detection of [Ne~V] in some individual AGN spectra is due to limited survey sensitivity rather than a deficit of coronal emission.
We find [Ne~V]/[Ne~III] is only weakly correlated with [O~III]/[O~II], in agreement with AGN photoionization models. Galaxies with high values of [Ne~V]/[Ne~III] for their [O~III]/[O~II] ratio have lower Eddington ratios (\edd) and are more likely to be classified as LIERs or Composites in the BPT diagram. The [Ne~V]/[O~III] ratio  shows a strong anticorrelation with [O~III] luminosity, with the most extreme coronal-line strengths found preferentially among the lowest-luminosity AGN. Together, these results provide evidence for a continuum between `low/hard' and `high/soft' accretion states, in analogy with stellar-mass black holes.
Comparison with a sample of $z=2$--$9$ [Ne~V]-detected galaxies suggests some, but not all, high-redshift AGN follow the same relations, offering a first test of whether this framework extends to the early universe.
\end{abstract}

%------------------------------------------------------------------------------------------------------
%                                                                                                    --
%                                          Section 1                                                 --
%                                                                                                    --
%------------------------------------------------------------------------------------------------------

\section{Introduction} \label{sec:intro} 

Active galactic nuclei (AGN) are powered by accretion onto supermassive black holes and play a central role in galaxy evolution across cosmic time.
At high redshift, however, they reside in low-metallicity, high-ionization environments where traditional narrow-line optical diagnostics fail.
Recent observations from the James Webb Space Telescope \citep[JWST][]{Gardner2006,Gardner2023} have revealed a significant population of AGN at high redshifts ($z \sim 3$--$11$), suggesting that supermassive black holes were already forming and influencing their environments in the early universe \citep[cf.][]{Jades_pop, epoch}.
These host galaxies are low-mass, metal-poor, and highly star-forming, posing particular challenges for traditional AGN identification methods \citep{highzagn, ChisholmNeV}.
In contrast, AGN at low redshift predominantly reside in massive \citep{kauffmann2003}, metal-rich galaxies \citep{Thomas2019} with low to moderate star formation rates \citep{Li2024_mm}.

Identifying these systems, therefore, requires methods that remain effective in low-metallicity, high-ionization, and high-redshift environments.
AGN can be identified through several complementary methods, including broad emission lines, radio emission, X-rays, mid-infrared colors, and optical forbidden lines \citep{Hickox}.
For high-$z$ galaxies, however, most of these methods begin to fail: broad emission lines can be hard to detect for low-mass black holes, and current radio, X-ray, and far-infrared surveys can only detect the very brightest high-$z$ objects.
Among these methods, optical narrow lines remain strong in high-$z$ galaxies. They are currently among the most accessible tools for AGN identification at high redshift, particularly for low-luminosity systems beyond the reach of current radio, X-ray, and mid-IR surveys.

The Baldwin, Phillips, and Terlevich (BPT) diagram \citep{OG_BPT} uses ratios of forbidden metal lines to permitted hydrogen lines to separate AGN from star-forming galaxies and remains in widespread use across optical spectroscopic surveys.
The BPT relies on the ratio of [O~III]~$\lambda$5008 over H$\beta$ compared to the ratio of [N~II]~$\lambda$6585 over H$\alpha$.
In this diagram, \Rthree~ acts as a proxy for ionization state, while [N~II]/H$\alpha$ acts as a proxy for metallicity.
As galaxies in the early universe are so metal-poor \citep{Isaac}, the [N~II]/H$\alpha$ ratio is significantly lower than what would be seen in AGN at low redshifts.
Thus, most high-$z$ AGN live in the high \Rthree, low \Ntwo\ region of the BPT diagram, in the same parameter space as low-metallicity star-forming galaxies, which also have high ionization parameters due to minimal line blanketing in stellar atmospheres (see Figure~8 from \citet{Jades_pop}).

Several alternative emission-line diagnostics have been proposed for use at high-$z$, including the Mass Excitation diagram \citep{MEx}, which replaces [N~II]/H$\alpha$ with stellar mass, and the OH-NO diagram \citep{OH-NO1, OH-NO2}, which uses [Ne~III]~$\lambda$3869/[O~II]~$\lambda\lambda$3727,3729 in its place.
However, these diagrams use \Rthree~ on the y-axis, and therefore they have the same issue: at high redshift, \Rthree~ is driven to extreme values by both low metallicity and high ionization parameter, which itself increases with redshift \citep[e.g.,][]{Cleri2026}, causing AGN and star-forming galaxies to occupy the same region of the diagram regardless of the x-axis diagnostic used.
Because \Rthree~is sensitive to both metallicity and ionization parameter, AGN identification at high redshift requires abundance-independent diagnostics \citep{Kewley2013, kewley2019, cleri2025}.
Coronal line emission offers one such alternative.

Coronal Lines (CLs) are forbidden emission lines with ionization potentials near or exceeding 100~eV, produced by highly ionized species in extreme environments.
In the extragalactic context, AGN are the dominant sources of CL emission: the CL luminosities of stellar sources such as supernova remnants, planetary nebulae, and Wolf--Rayet stars fall many orders of magnitude below those of AGN, and tens of thousands of such sources would be needed to produce comparable emission \citep{CLR}.
CLs are therefore considered robust AGN tracers \citep{Gilli2010,CLeri2023,Cleri_MEx,Negus,CLR}.
Among the brightest and most commonly detected CLs are [Ne~V]~$\lambda\lambda$3347,3427, [Fe~X]~$\lambda$6374, and [Fe~VII]~$\lambda\lambda$5720,6087 \citep{CLASS}.
One practical limitation is that CLs are intrinsically faint relative to other AGN emission lines, with fluxes typically a factor of 2 to over 3 orders of magnitude weaker than [O~III]~$\lambda$5008 and H$\alpha$ \citep{CLASS}, limiting their detectability even in well-observed samples.

Of the available CLs, [Ne~V]~$\lambda\lambda$3347,3427 (ionization potential 97.1~eV) offers the best opportunity for a statistical study.
It is the brightest optical CL and the most frequently detected in optical spectroscopic surveys \citep{CLASS}.
In addition, as a noble gas, neon is not depleted onto dust grains, making [Ne~V] a more reliable tracer than iron CLs in dusty environments, in spite of its comparatively blue wavelength \citep{Negus,NeV_Dust}. 

The geometry of the CLR adds another layer to this picture.
Reverberation mapping places the [Ne~V]~$\lambda$3427-emitting region of the quasar COS168 (SDSS J095910.30+020732.2) at a rest-frame lag of $\sim$281.7~light-days \citep{clr_location}, near or beyond the characteristic torus radius of $\sim$297~light-days and well beyond the dust sublimation radius of $\sim$143~light-days.
The CLR is therefore embedded in one of the dustier environments of the AGN.
Because neon is not grain-depleted, dust does not suppress [Ne~V] the way it suppresses iron CLs.
The strength of [Ne~V] relative to lower-ionization lines therefore reflects the geometry of the CLR and the local ionizing conditions.
At high accretion rates, radiation pressure may further expand the dust sublimation radius, altering the geometry of the CLR and modifying the relative flux of $>$97~eV photons compared to that of the lower-ionization lines, which are produced further out.

[Ne~V] has not yet been used for AGN demographic studies: existing catalogs contain too few detections for a statistical assessment of which AGN produce CL emission and which do not.
The six most extensive CL catalogs, from largest to smallest, are as follows.
The largest comprises 1,215 [Ne~V] AGN from the Baryon Oscillation Spectroscopic Survey (BOSS), part of the Sloan Digital Sky Survey III (SDSS-III), Data Release 12 \citep{BOSS, DR12} processed by the Portsmouth group\footnote{\url{https://www.sdss4.org/dr12/spectro/galaxy_portsmouth/}} as described in \citet{feuillet}.
The remaining five are: 258 AGN with any of 20 different CLs in the Coronal Line Activity Spectroscopic Survey \citep[CLASS;][]{CLASS} (SDSS-III DR8); 88 [Ne~V] AGN in the Lockman Hole Field \citep{Li2024}; 81 [Fe~X] detections in dwarf galaxies from SDSS-I \citep{molina}; 71 galaxies in MaNGA \citep{Negus}; and 25 high-$z$ objects observed with Hubble Space Telescope \citep{Cleri_MEx}.
In absolute terms, these detections are few in number, and the [Ne~V]-selected AGN represent only a fraction of the AGN in their parent samples.; for example, the \citet{feuillet} sample contains 1,215 [Ne~V] detections out of 2,847 BPT AGN, a detection fraction of 42.7\%.
The median [OIII] 5008 / [Ne V] 3427 flux ratio among detected objects is approximately 10-20, with some dependence on signal-to-noise (see \S\ref{sec:Data}), limiting its detectability even in well-observed samples.
This suggests that some AGN that lack [Ne~V] are non-detected due to low S/N.
However, it has been demonstrated that some luminous AGN do not show CL emission \citep{NCL_real}.

If we are to use \NeV~as an AGN indicator in the high-redshift universe, its selection effects must first be characterized at low redshift, where the physical conditions of CL-emitting AGN can be studied in detail, and the populations [Ne~V] selects or omits can be identified.
Taken together, existing catalogs contain too few objects to statistically assess which AGN produce CL emission and which do not.

\begin{figure}[ht]
\plotone{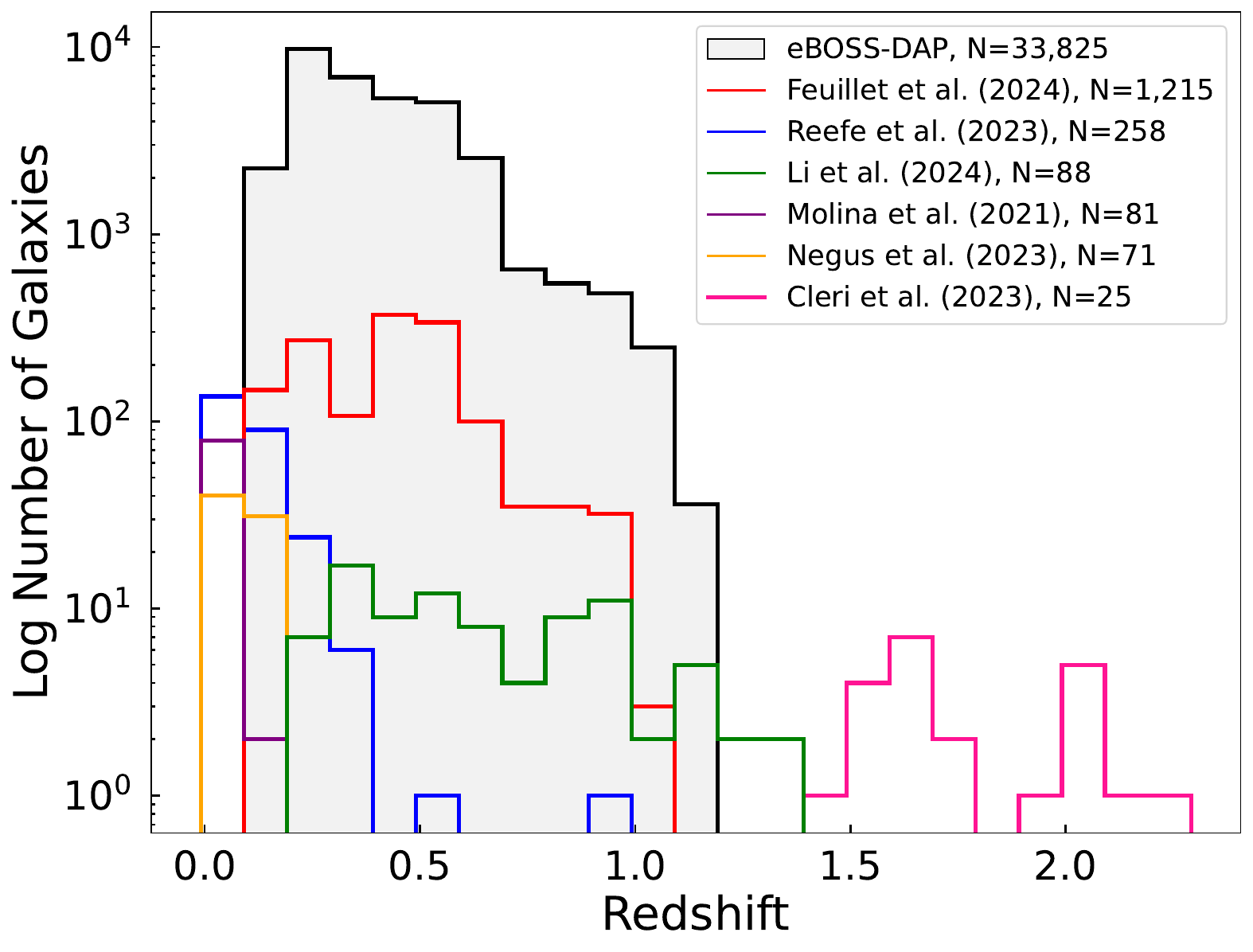}
\caption{Redshift distribution of [Ne V]-detected galaxies in the eBOSS-DAP (gray, N = \numNeVgal, this work) compared to the six largest CL catalogs in the literature, shown on a logarithmic scale: \citet{feuillet} (N = 1,215 [Ne~V] AGN from BOSS (SDSS-III) DR12, red); \citet{CLASS} (N = 258 AGN with any of 20 CLs, SDSS-III DR8, blue); \citet{Li2024} (N = 88 [Ne~V] AGN in the Lockman Hole Field, green); \citet{molina} (N = 81 [Fe~X] detections in dwarf galaxies from SDSS-I, purple); \citet{Negus} (N = 71 galaxies in MaNGA, orange); and \citet{Cleri_MEx} (N = 25 high-$z$ objects observed with HST, pink).
The eBOSS-DAP sample exceeds the combined total of all six literature catalogs (1,738 objects) by more than an order of magnitude at every redshift bin where they overlap, and extends continuously to $z \sim 1.12$.
\label{fig:lit_comp}}
\end{figure}

The Sloan Digital Sky Survey IV (SDSS-IV) Extended Baryon Oscillation Spectroscopic Survey \citep[eBOSS;][]{eBOSS} represents a largely untapped resource for this purpose: its target selection produces a high AGN fraction, and the newly released eBOSS - Data Analysis Pipeline \citep[eBOSS-DAP;][]{eBOSS-DAP} yields the largest sample of CL-emitting galaxies assembled to date.
The literature currently contains approximately 1,738 CL-emitting objects across all published catalogs (see Figure~\ref{fig:lit_comp} in \S\ref{sec:Data}); eBOSS exceeds this total by more than an order of magnitude, enabling the first statistical assessment of which AGN produce coronal emission and which do not.

This paper characterizes the physical conditions that give rise to CL emission and refines \NeV~as a statistical AGN diagnostic drawing from a sample of \numNeVgal~[Ne~V]-detected galaxies from the eBOSS-DAP.
An outline of the paper is as follows:
In \hyperref[sec:Data]{Section 2}, we describe the eBOSS-DAP catalog and define the subsamples used in the analysis.
In \hyperref[sec:methods_extra]{Section 3}, we describe the star-forming galaxy matching, spectral stacking, and line-correction procedures used throughout the paper.
In \hyperref[sec:nev-ncl]{Section 4}, we examine the distribution of \NeV~detections across the BPT diagram and we  define two subsamples for comparison based on their [Ne\,V] signal-to-noise (S/N): NeV-AGN and non-coronal AGN (NC-AGN).
In \hyperref[sec:pop_comp]{Section 5}, we compare the physical properties of NeV-AGN and NC-AGN and stack NC-AGN spectra to test whether their [Ne\,V] non-detections reflect a severe deficit of coronal emission or a survey sensitivity limitation.
In \hyperref[sec:ner_or]{Section 6}, we investigate the shape of the AGN ionizing continuum using the [Ne~V]/[Ne~III]--[O~III]/[O~II] diagram.
In \hyperref[sec:nev_oiii_loiii]{Section 7}, we test how [Ne~V]/[O~III] depends on AGN luminosity, including a comparison to high-redshift AGN.
In \hyperref[sec:heii]{Section 8}, we compare \NeV~detection to the He~II~$\lambda$4686/H$\beta$ diagnostic and show that the two selection methods identify substantially different AGN populations.
Finally, we provide a summary of our main findings in \hyperref[sec:summary]{Section 9}.

Throughout this paper, we adopt a flat $\Lambda$CDM cosmology from \citet{Planck2018}; where an initial mass function is required (e.g., for stellar masses and star formation rates, \S\ref{sec:Data}), we assume a \citet{chabrier_IMF} IMF. For [Ne~V], we use only the stronger (3427~\AA) line in all line-ratio quantities throughout this paper, reserving the co-added \NeVp~doublet flux for S/N thresholds and equivalent widths.

%------------------------------------------------------------------------------------------------------
%                                                                                                    --
%                                          Section 2                                                 --
%                                                                                                    --
%------------------------------------------------------------------------------------------------------

\section{Data \& Models} \label{sec:Data} 

We utilize optical spectra from the Sloan Digital Sky Survey \citep{York:2000} Data Release 17 \citep[SDSS DR17;][]{Abdurrouf2022} obtained as part of the BOSS \citep{Eisenstein2011} and eBOSS \citep{eBOSS} surveys. (Hereafter, we refer to the dataset collectively as eBOSS).
eBOSS targeted galaxies at $z \lesssim1$ using a variety of color-based photometric selection algorithms, producing a sample of \nsample~spectra of Luminous Red Galaxies, Emission Line Galaxies, and compact blue galaxies initially targeted as quasars \citep{eBOSS}; \nsampledap~of these were successfully processed by the eBOSS-DAP and form the working sample used throughout this paper.
Spectra were obtained through 2\arcsec\ diameter fibers using the SDSS-BOSS spectrographs \citep{Smee:2013}, covering 3610--10140~\AA\ at $R = 1560$--$2650$.

\begin{figure*}[ht]
\plotone{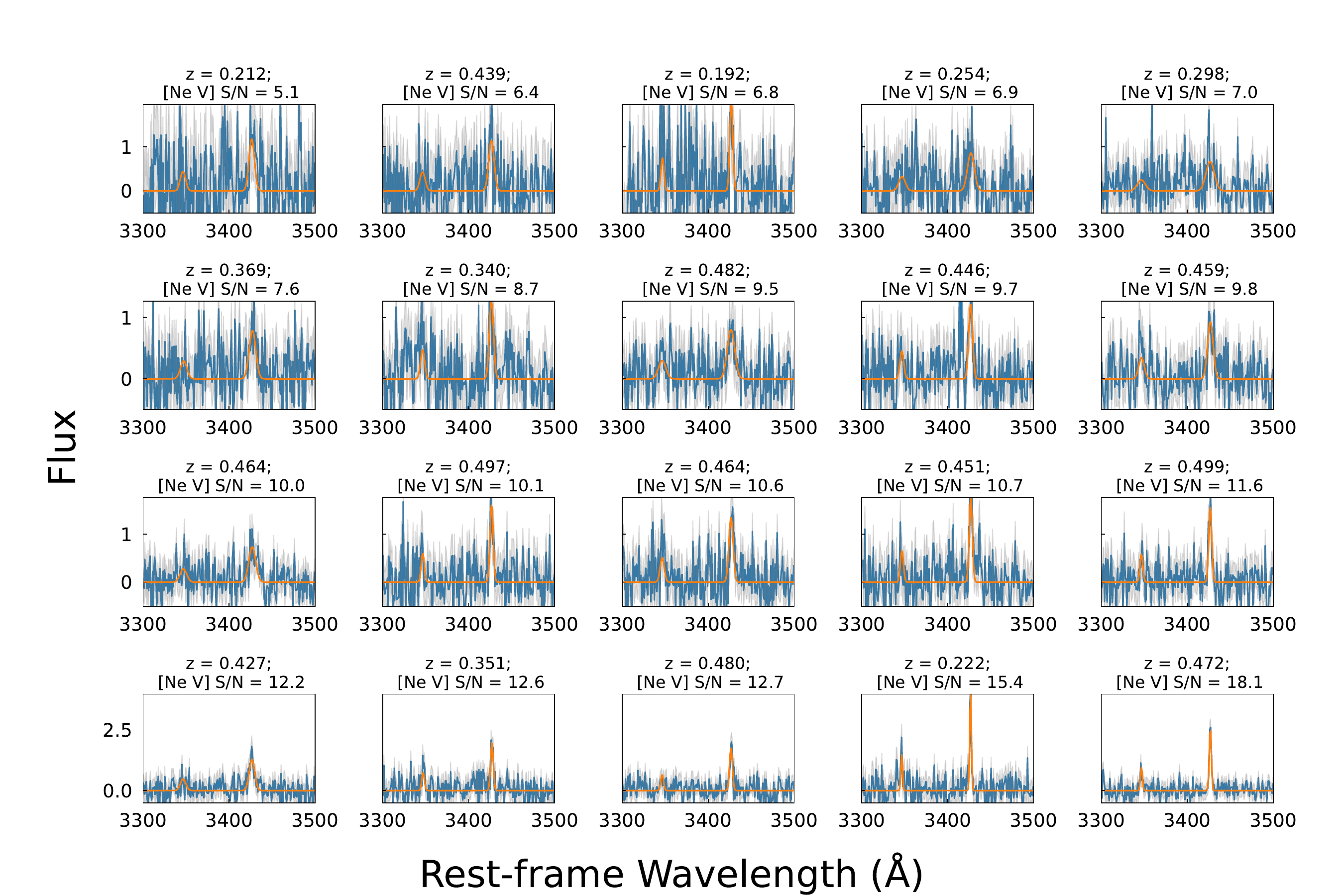}
\caption{Twenty randomly selected spectra drawn from the BPT-classified AGN in our sample, sorted by increasing [Ne~V] S/N, illustrating the range of signal-to-noise ratios observed in the eBOSS-DAP.
Spectra are selected to have \NeVp~S/N $\geq$ 5 and redshifts $0.147 < z < 0.55$.
The [Ne~V] S/N is computed from the co-added doublet flux; tying the two lines kinematically in the eBOSS-DAP recovers a higher combined S/N than the stronger line alone, maximizing detection power for this intrinsically faint doublet.
Each panel shows the rest-frame wavelength region surrounding the [Ne~V] doublet, with the measured redshift and [Ne~V] S/N given above each panel.
\label{fig:NeV_grid}}
\end{figure*}

We processed the spectra using the eBOSS Data Analysis Pipeline \citep[eBOSS-DAP;][]{eBOSS-DAP}, which builds on the MaNGA-DAP \citep{MaNGA-DAP} framework to deliver uniform emission-line fluxes, absorption indices, and stellar kinematics for the full sample.
The eBOSS-DAP improves the detectability of faint doublets such as [Ne~V]~$\lambda\lambda$3347,3427 by tying the kinematics of weak lines to those of stronger nearby features and fixing doublet flux ratios to their quantum-mechanically prescribed values; for [Ne~V], the flux ratio [Ne~V]~3347/[Ne~V]~3427 is fixed at 0.366.
The catalog and source code are publicly available; see the Data Availability statement at the end of this paper for details.

Accretion rate and black hole mass are likely to play a role in determining whether an AGN produces detectable CL emission.
We characterize these using the stellar velocity dispersion $\sigma_{*}$, measured from the stellar continuum by the eBOSS-DAP, and the Eddington ratio $L_{\rm Bol}/L_{\rm Edd} = $\edd.
Black hole mass is estimated by converting $\sigma_{*}$ via the M$_{\rm BH}$--$\sigma_{*}$ relation \citep{Msig}. 
The Eddington ratio requires an estimate of AGN bolometric luminosity, which we obtain from $L_{\rm [OIII]}$ in two steps.
First, we correct $L_{\rm [OIII]}$ for dust attenuation using the Balmer-decrement $E(B-V) = 1.97 \log_{10}[(\mathrm{H}\alpha/\mathrm{H}\beta)/3.06]$ and the \citet{Calzetti} attenuation law, adopting $k_{\rm OIII} = 4.463$ at 5008~\AA; we use an intrinsic $\mathrm{H}\alpha/\mathrm{H}\beta = 3.06$, the AGN-model-derived value described in \S\ref{sec:methods_extra}, rather than the standard case-B value of 2.86, which our own models show does not hold for narrow-line-region gas (\S\ref{sec:methods_extra}).
We then convert the dust-corrected $L_{\rm [OIII]}$ to a bolometric luminosity using the empirical correction factor of 142 from \citet{Lamastra2009}.
The Eddington luminosity defined as $L_{\rm Edd} = 1.26 \times 10^{38} (M_{\rm BH}/M_{\odot})$~erg~s$^{-1}$ \citep{Ledd}.
Wherever $\sigma_{*}$ appears, we apply additional quality cuts of $\sigma_{*} < 449~\mathrm{km~s}^{-1}$ and $\sigma_{*}$~uncertainty~$< 50~\mathrm{km~s}^{-1}$.

We use the \drake~Value-Added Catalog (VAC), which contains stellar masses, star formation rates, and Legacy Survey \citep{Dey:2019} photometry.
Stellar masses and star formation rates were measured by fitting spectral energy distributions (SEDs) with CIGALE \citep{CIGALE}, using Legacy Survey $gri$ photometry \citep{Dey:2019}, WISE W1-W3 photometry \citep{Wright:2010}, and the D$_n$4000 index from the eBOSS-DAP as inputs.
The SED models assume a delayed $\tau$ star-formation history with a burst component, \citet{sps} stellar population templates with a Chabrier IMF \citep{chabrier_IMF}, and dust attenuation and emission laws from \citet{Calzetti} and \citet{Draine}, respectively.
Typical stellar mass uncertainties are $\pm$0.07~dex.

%---------------------------------------------------------------------------------------------------
%                                  Subsection 2.1                                                 --
%---------------------------------------------------------------------------------------------------

\subsection{The [Ne\,V] Parent Sample}\label{subsec:nev_parent}
From the full eBOSS-DAP sample of \nsampledap~spectra, we select galaxies with $z > 0.147$ and \NeVp~S/N $>5$. The redshift limit follows the recommendation of \citet{eBOSS-DAP} to exclude data in which the [Ne~V] doublet falls too close to the blue edge of the spectral window.
We adopt S/N $> 5$ as the detection threshold for the co-added [Ne~V] doublet, rather than the S/N $> 3$ typically applied to stronger optical lines, following the eBOSS-DAP recommendation to raise the S/N limit for lines that are both blue and intrinsically weak, conditions that [Ne~V] satisfies on both counts.
Application of these criteria yields \numNeV~spectra (\numNeVgal~unique galaxies) with [Ne~V], which exceeds the combined literature total of 1,738 by more than an order of magnitude. The redshift distribution of our sample and the samples drawn from the literature is shown in Figure~\ref{fig:lit_comp}.
In Figure~\ref{fig:NeV_grid}, we show the fitted [Ne~V] lines for a random selection of eBOSS spectra.

\begin{figure}[ht]
\plotone{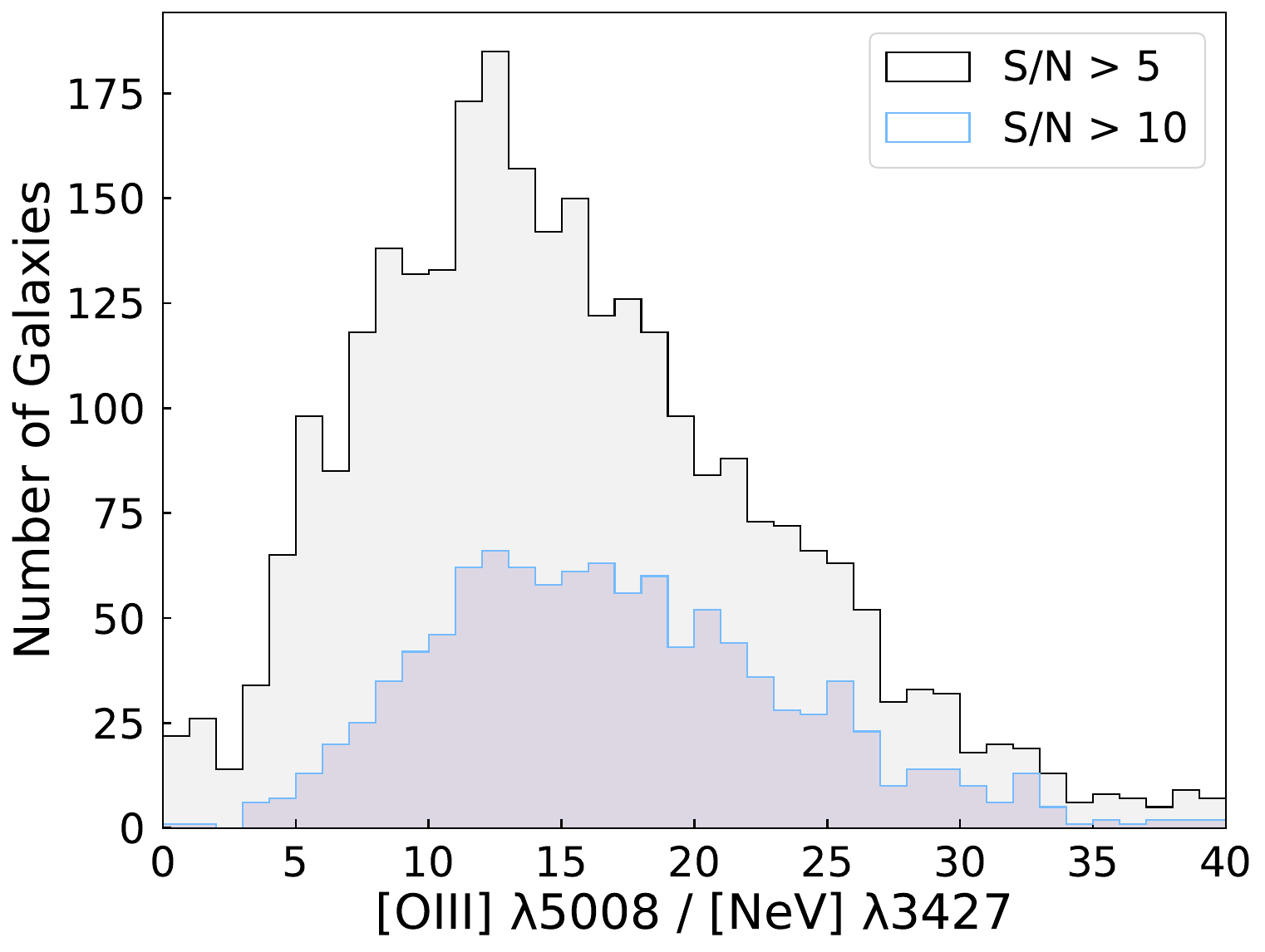}
\caption{The ratio of [O~III]~5008 to [Ne~V]~3427 flux for BPT AGN with $z > 0.147$, shown for two detection thresholds: [O~III]~and [Ne~V]~3427 S/N $\geq$ 5 (black, N = 2,878) and S/N $\geq$ 10 (blue, N = 1,064).
Both distributions peak near a ratio of 13.5; the S/N $\geq$ 5 sample has a median of 14.5 and an interquartile range of 9.9 to 20.3, while the S/N $\geq$ 10 sample has a median of 16.5 and an interquartile range of 12.1 to 21.4.
This ratio quantifies the intrinsic faintness of [Ne~V] relative to [O~III], which limits [Ne~V] detectability even in well-observed samples.
\label{fig:OIII_NeV}}
\end{figure}

Despite being among the brightest coronal lines, [Ne~V] is comparatively weak relative to \OIII, a common AGN bolometric indicator \citep{CLASS}. Figure~\ref{fig:OIII_NeV}, shows a histogram of the [O~III]/[Ne~V] line ratio for BPT-selected AGN. [Ne~V]~3427 has a median [O~III]/[Ne~V] flux ratio of 14.5 among detected objects, with an interquartile range of 9.9 to 20.3.
This suggests that, even where the doublet-tying strategy substantially boosts the effective S/N, \NeV~remains intrinsically faint enough that detection in the eBOSS-DAP sample is a challenge regardless of methodology.
The breadth of this ratio, spanning roughly a factor of two between the 25th and 75th percentiles, further suggests that a single characteristic [O~III]/[Ne~V] ratio does not adequately describe the sample; we return to this diversity in ionizing conditions in \S\ref{sec:ner_or}.

%---------------------------------------------------------------------------------------------------
%                                  Subsection 2.2                                                 --
%---------------------------------------------------------------------------------------------------
\subsection{NeV-AGN and NC-AGN Samples} \label{subsec:nev-ncl_data}
From our parent sample, we define two subsamples of AGN: one where [Ne V] is detected (hereafter `NeV-AGN'), and one where [Ne V] is not detected.  We refer to the latter as `non-coronal AGN' (hereafter 'NC-AGN'), although we recognize that coronal emission may be present but too weak to detect in these galaxies.
The NeV-AGN and NC-AGN samples are defined to enable a direct comparison of the physical properties that distinguish AGN with and without detectable [Ne~V] emission.

To select these subsamples, we first apply a cut of S/N $>$ 3 for \Hb, \OIII, \Ha, and \NII.
With this subset, we then retain only the spectra classified as AGN by the \citet{LawBPT} BPT diagram.  This classification excludes LIERs and Composites. 
Once only AGN remain, we trim to those within the redshift range $0.147 < z < 0.55$.
This redshift cut follows the recommendation of \S2.4 of \citet{eBOSS-DAP} to omit data in which lines of interest fall too close to the edges of the spectra (for this paper, \NeV~on the low end and \Ha~on the high end).
We additionally require SPECPRIMARY~$=1$ to exclude duplicate observations, bringing the 9,419 spectra to a core sample of 8,759 BPT-classified AGN.

\begin{figure*}[ht]
\plotone{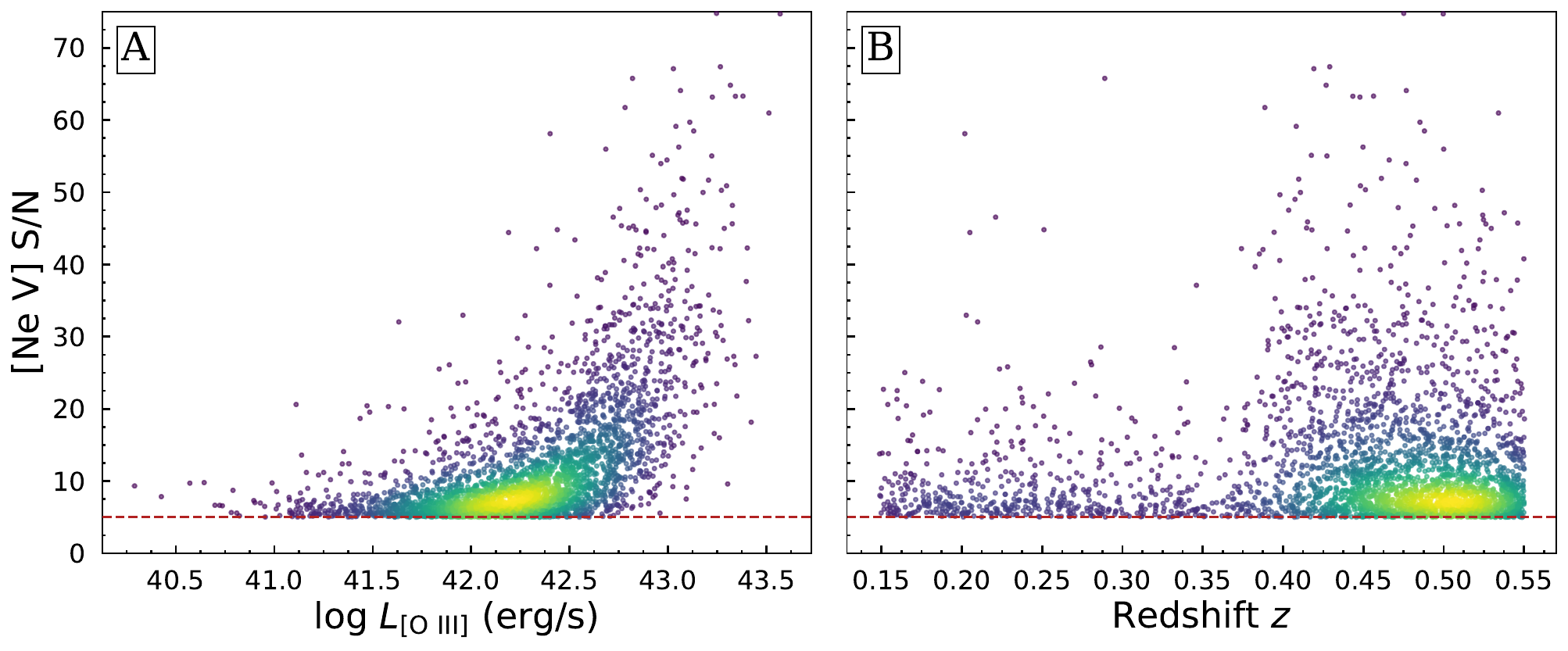}
\caption{[Ne~V] signal-to-noise ratio (S/N) as a function of \textbf{(A)} \OIII\ luminosity and \textbf{(B)} redshift, for the NeV-AGN sample (\S\ref{subsec:nev-ncl_data}; N = 3,088), colored by local point density. The dashed line marks the S/N $= 5$ detection threshold used throughout this work.
\label{fig:NeVSNR_vs_LOIII_z}}
\end{figure*}

We define NeV-AGN as BPT AGN spectra meeting a \NeVp~detection threshold of S/N $>$ 5, yielding 3,088 spectra.
We construct the NC-AGN sample from BPT AGN with \NeVp~S/N $<$ 3, yielding 2,626 spectra.
While [Ne~V] is intrinsically a weak line (\S\ref{sec:intro}), Figure~\ref{fig:NeVSNR_vs_LOIII_z} shows that the NeV-AGN sample includes a substantial tail of high-significance detections, with S/N reaching nearly 90 for the most luminous, lowest-redshift AGN.

In \S\ref{sec:nev-ncl}, we will compare the physical properties of the NC-AGN population to those of the NeV-AGN to characterize the conditions that give rise to coronal line emission.

%---------------------------------------------------------------------------------------------------
%                                  Subsection 2.3                                                 --
%---------------------------------------------------------------------------------------------------

\subsection{Ne-O Sample} \label{subsec:ne-o_data}
We also define a Ne-O subsample which we use in \S\ref{sec:ner_or} to analyze extreme UV SEDs of our AGNs via their [Ne~V]/[Ne~III] and [O~III]/[O~II] line ratios.
Unlike the previous subsample, these galaxies are not limited to those classified as AGN by a BPT diagram; instead, they are classified as AGN solely on the basis of the presence of coronal lines.
To create this sample, we cut at S/N $>$ 5 for \OII, \OIII, \NeIII, and \NeVp.
We further restrict to the redshift range $0.147 < z < 0.55$ and \texttt{SPECPRIMARY = 1}, as done for the NeV-AGN and NC-AGN samples (\S\ref{subsec:nev-ncl_data}).
This yields a sample of 2,458 galaxies exhibiting strong NeV emission and well-measured oxygen lines.

%------------------------------------Subsubsection 2.3.1------------------------------------

\subsubsection{AGN Photoionization and Stellar Population Models} \label{subsubsec:agn_models}
To compare against the Ne-O sample in \S\ref{sec:ner_or}, we use a selection of the photoionization model library presented in \citet{cleri2025}.
These models have two prescriptions for ionizing sources.
The AGN models are computed using XSPEC-based spectral energy distributions that prescribe a blackbody accretion disk with a Compton upscattering component reproducing the soft X-ray excess commonly observed in local AGN.  Black hole mass is varied over $\log M_\mathrm{BH}/M_\odot \in \{3, 4, 5, 6, 7, 8, 9\}$, and ionization parameters range from $-4 \leq \log U \leq -1$ in steps of 0.25~dex.
Gas-phase metallicity ($Z_{\rm gas}$)  varies between $0.001~Z_{\odot}$ and $2.0~Z_{\odot}$ in 13 steps assuming a solar relative abundance pattern, and hydrogen density is varied over $\log (n_{\rm H}/{\rm cm}^{-3}) \in \{2, 3, 4\}$. The Eddington ratio is held constant for these models at $\log(\lambda_{\rm Edd}) = -1$, as any variations in \edd\ are degenerate with variations in $M_\mathrm{BH}$.

The stellar population models are drawn from the Binary Population and Spectral Synthesis library \citep[\textsc{BPASS} v2.2.1;][]{Stanway2018}, spanning ages $6.0 \leq \log(\mathrm{age/yr}) \leq 11.0$, metallicities $10^{-5} \leq Z_\star \leq 0.040$, a range of initial mass functions, and both single-star and binary-star formation channels.
Both the AGN photoionization models and the \textsc{BPASS} stellar population models are computed using \textsc{Cloudy} \citep[version C23.01;][]{Ferland2017, Gunasekera2023} over the same grid of ionization parameters.

\begin{figure}[ht]
\plotone{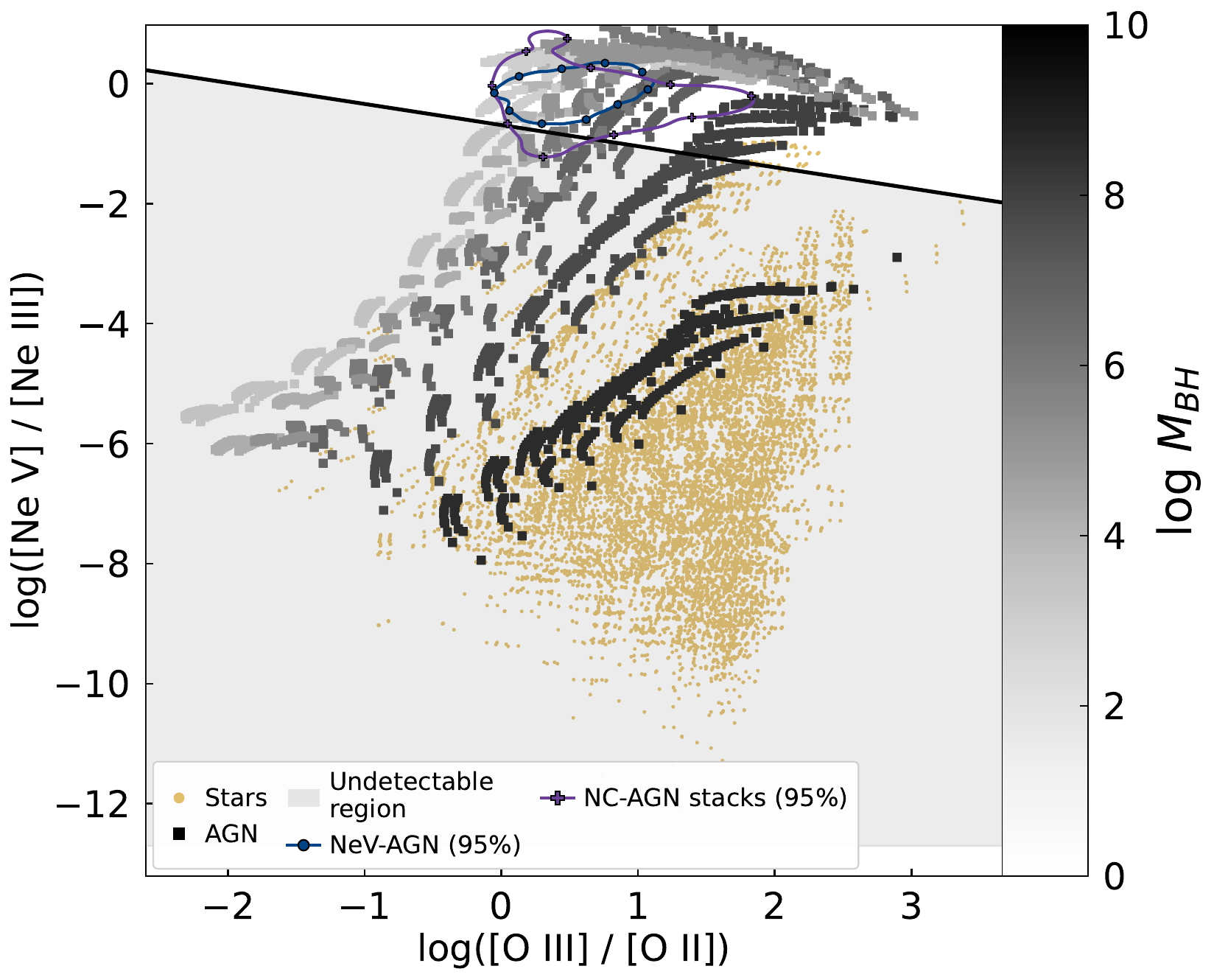}
\caption{\NeR\ vs.\ \OR\ for the \citet{cleri2025} photoionization models at solar metallicity, showing both AGN and stellar photoionization models, together with three empirical comparisons.
AGN models (gray--black squares) are color-coded by black hole mass.
Stellar models from \textsc{bpass} \citep{Stanway2018} are shown as gold points and populate a broad, diffuse cloud confined to low log ([Ne V] / [Ne III])~values, reflecting the inability of stellar photospheres to produce the $>97$~eV photons required to generate [Ne~V]~emission.
The blue contour marks the 95\% extent of the NeV-AGN sample and the purple contour marks the 95\% extent of the NC-AGN stacks (\S\ref{subsec:ncl-stacking}); both samples, along with the solid detection-limit line, are computed using the star-forming-corrected \OIIcorr, \OIIIcorr, and \NeIIIcorr\ quantities described in \S\ref{sec:methods_extra}.
The shaded region below the solid line denotes eBOSS-survey-inaccessible parameter space at [Ne~V]~S/N~$=5$.
The clear separation between the AGN model tracks and the stellar model locus in the log ([Ne V] / [Ne III])~dimension is consistent with the conclusion that the observed CL emission in eBOSS cannot be reproduced by stellar populations alone.
\label{fig:ne53_o32_full_grid}}
\end{figure}

In Figure~\ref{fig:ne53_o32_full_grid} we show a preliminary comparison between the \citet{cleri2025} photoionization models and the eBOSS data using the \NeR\ and \OR~line ratios.   The stellar and AGN models have comparable [O~III]/[O~II] line ratios but very different [Ne~V]/[Ne~III] values (with the exception of AGN hosted by $\sim10^{10}$~M$_{\sun}$ black holes.) This highlights the utility of [Ne V] as an AGN diagnostic.  
However, as noted above, [Ne V] is a comparatively weak line; the solid line marks the survey's 5th-percentile [Ne~V] detection limit as a function of ionization state.
The 95\% contour is shown for the NeV-AGN (blue solid line) and the NC-AGN stack sample (purple; \S\ref{subsec:ncl-stacking}), where line fluxes were measured from stacked eBOSS spectra to increase the detection limit.
All empirical line ratios use star-formation-corrected [O~II], [O~III], and [Ne~III] fluxes, following the procedure described in \S\ref{sec:methods_extra}.
 Clearly, within the sensitivity range of the eBOSS survey, we do not expect to detect [Ne~V] emission associated with star formation.
We explore the data-model comparison further in \S\ref{sec:ner_or}.

%------------------------------------------------------------------------------------------------------
%                                                                                                    --
%                                          Section 3                                                 --
%                                                                                                    --
%------------------------------------------------------------------------------------------------------
\section{Star Formation Corrections and Stacking Methodology} \label{sec:methods_extra}

Several analyses in this paper rely on procedures applied to the NeV-AGN and NC-AGN samples: matching each galaxy to a pool of star-forming galaxies (SFGs) of similar redshift, stellar mass, and color; stacking the matched SFGs (and, elsewhere in this paper, matched AGN) to recover signal below the per-object detection threshold; and using the resulting SFG stacks to correct \OII, \OIII, and \NeIII\ for star-forming contamination.
We describe each of these three steps below.

%---------------------------------------------------------------------------------------------------
%                                  Subsection 3.1                                                 --
%---------------------------------------------------------------------------------------------------

\subsection{Stacking Procedure} \label{subsec:stacking_methodology}

Because \NeV~and other weak emission lines often fall below the detection threshold in individual spectra, we combine multiple spectra into a single higher signal-to-noise spectrum to recover weak lines.
We use the same median-stacking procedure throughout this paper, whether stacking matched SFGs to build a per-galaxy correction, stacking galaxies on a grid of BPT-diagram position (\S\ref{sec:nev-ncl}) or a grid of black hole mass and \OIII~luminosity (\S\ref{subsec:ncl-stacking}). 
For each stack, we deredshift each contributing spectrum, correct for Milky Way extinction, and interpolate onto a common rest-frame wavelength grid.
We renormalize each spectrum to a shared flux scale using the median flux in the 3490--3540~\AA~window (which is devoid of strong spectral features), then combine via the pixel-wise median.
Uncertainties are estimated from the median absolute deviation (MAD) of the contributing spectra at each pixel; MAD is scaled by a factor of 1.4826, which converts it to an equivalent standard deviation for a normal distribution, and the result is divided by the square root of the number of contributing spectra.
The stacked spectrum is then fit with the same emission-line fitting pipeline used for individual eBOSS spectra \citep{eBOSS-DAP}, yielding stacked line fluxes, equivalent widths, and their uncertainties.

\begin{figure}[ht]
\plotone{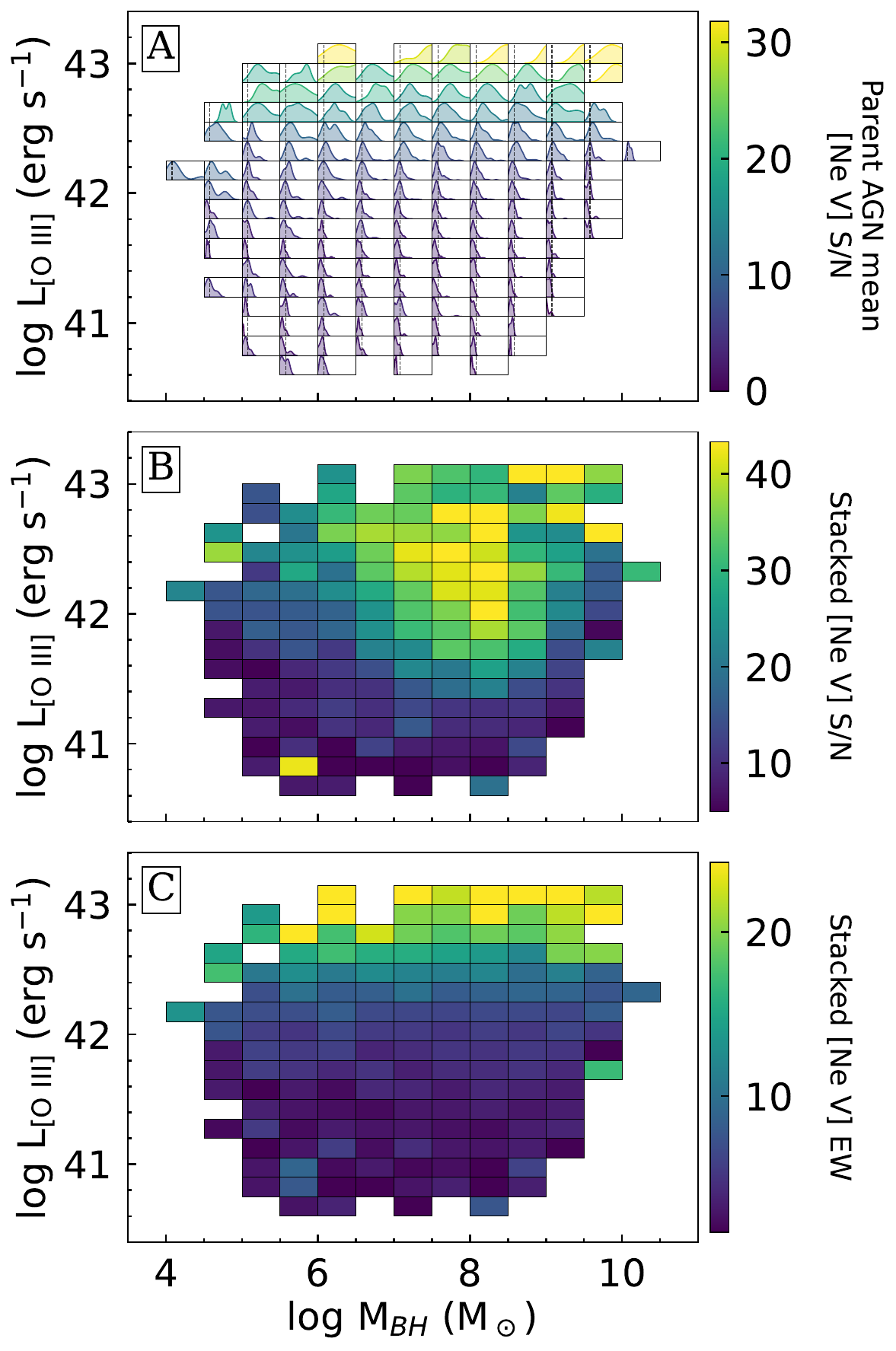}
\caption{AGN stacked on a grid of black hole mass and \OIII~luminosity.
\textbf{Panel A}: 1-D Kernel Density Estimates (KDEs) of the [Ne~V] S/N distribution in each bin, with the x-axis spanning 0.0 $<$ [Ne~V] S/N $<$ 22.95; bins are color-coded by mean [Ne~V] S/N, and a dashed vertical line marks S/N $= 5$, the detection threshold adopted in this work.
\textbf{Panel B:} [Ne~V] S/N of the stacked spectrum in each bin. The S/N depends on both the intrinsic [Ne\,V] strength and the number of AGN available to stack in each bin. 
\textbf{Panel C:} [Ne~V] equivalent width of the stacked spectrum in each bin. 
Bins containing fewer than five galaxies are omitted.
[Ne~V] is detected in 150 of the 162 bins (92.6\%) with at least 5 galaxies per stack, reaching 100\% detection for bins with at least 15 galaxies.
\label{fig:mbh-loiii-agn-grid}}
\end{figure}

As one application of this procedure, we stack AGN spectra on a grid of black hole mass and \OIII~luminosity, with no restriction on [Ne\,V]~detection, to test how [Ne\,V]~strength depends on these properties across the full AGN population.
We draw this parent sample from the full BPT-classified AGN population ($0.147 < z < 0.55$; \S\ref{subsec:nev-ncl_data}), restricted to sources with a well-measured stellar velocity dispersion ($\sigma_{*} < 449~\mathrm{km~s^{-1}}$, $\sigma_{*}$ uncertainty $< 50~\mathrm{km~s^{-1}}$).
For each galaxy we estimate $M_{\rm BH}$ from $\sigma_{*}$ via the M--$\sigma$ relation of \citet{Msig}, and bin galaxies on a grid of $0.55$~dex in $\log M_{\rm BH}$ and $0.15$~dex in $\log L_{\rm [OIII]}$; the wider $M_{\rm BH}$ step reflects the steep slope of the M--$\sigma$ relation ($M_{\rm BH} \propto \sigma_{*}^{5.64}$).
Bins containing fewer than five galaxies are excluded; there is no upper cap on bin size.
Figure~\ref{fig:mbh-loiii-agn-grid} shows the results of this stacking procedure across the black hole mass--\OIII~luminosity plane. We use the resulting data in our analysis of the [Ne\,V]/[O\,III]--$L_{[O\,III]}$ relation in \S\ref{sec:nev_oiii_loiii}. We construct an analogous set of stacks restricted to NC-AGN in \S\ref{subsec:ncl-stacking} and use these data in Figures \ref{fig:nev-ncl-host-corner} and \ref{fig:nev-ncl-agn-corner} to explore the properties of AGN with [Ne\,V] below the detection limit in individual eBOSS spectra.

%---------------------------------------------------------------------------------------------------
%                                  Subsection 3.2                                                 --
%---------------------------------------------------------------------------------------------------

\subsection{Star-Forming Galaxy Matching} \label{subsec:sfg_matching}

The 2\arcsec\ diameter eBOSS fibers capture $\sim30-80$\% of the galaxies' total light and thus sample gas ionized by both the AGN and star formation within the fiber aperture.
As a result, \OII, \OIII, and \NeIII, all of which can be produced by star-forming regions as well as by AGN, require a correction for this star-forming contribution before they can be used as tracers of AGN ionization conditions alone.
To correct [O II], [O III], and [Ne III] for contamination from star formation, we match each galaxy in the full NeV-AGN and NC-AGN samples to a pool of star-forming galaxies (SFGs) of similar redshift, stellar mass, and color.
The SFG pool is drawn from the \citet{LawBPT} star-forming region of the BPT diagram using the same \Ha, \Hb, \NII, and \OIII~S/N $>$ 3 cuts applied to the AGN sample. We also require the galaxies to have Legacy Survey $g-r$ colors, and we remove duplicate spectra by requiring SPECPRIMARY~$=1$. No additional line S/N cut is imposed on individual SFGs, since the stacking boosts the signal of weaker lines like [Ne~III].
For each galaxy, we require at least three SFG matches within $\Delta z \leq 0.02$, $\Delta \log M_{*} \leq 0.2$~dex, and $\Delta(g{-}r) \leq 0.05$~mag.
This matching is applied identically to the full NeV-AGN and NC-AGN samples (\S\ref{subsec:nev-ncl_data}) and the Ne-O sample used in \S\ref{sec:ner_or}, which is not restricted to BPT AGN.

Extending this procedure to the BPT-grid stacked bins described in \S\ref{subsec:stacking_methodology} requires an additional step, since each bin combines many AGN spanning a range of redshift, stellar mass, and color. We therefore apply the per-galaxy matching criteria above to every AGN in a given bin individually, and take the union of all resulting SFG matches, discarding duplicates, to build a SFG sample for that bin. This sample is then combined into one SFG stack per bin following the procedure used in \S\ref{subsec:stacking_methodology}. This SFG matching procedure succeeds for all 71 BPT-grid bins (\S\ref{sec:nev-ncl}), with a median of 4,092 SFGs per bin. The M$_{\rm BH}$--[O~III]~luminosity stacks used in \S\ref{subsec:ncl-stacking} are not SFG-corrected.

Having established SFG matches for each galaxy, we now use the resulting stacks to correct [O~II], [O~III], and [Ne~III]~for star-forming contamination.
For each galaxy with a valid SFG stack, we subtract the stacked SFG flux from the galaxy's own [O~II], [O~III], and [Ne~III]~flux to produce [O~II]$_{\rm corr}$, [O~III]$_{\rm corr}$, and [Ne~III]$_{\rm corr}$, with uncertainties added in quadrature.
In galaxies where the SFG flux matched or exceeded the AGN flux in any of the three lines, the galaxy is dropped from the corrected sample, since \ORcorrs\ and \NeRcorrs\ both require all three lines to be usable.
Among the BPT-position stacks, the masked percentage is 36.6\%, 35.2\%, and 33.8\% for [O~II], [O~III], and [Ne~III], respectively; this is driven almost entirely by LIER and Composite stacks (masked at 55--60\% and 50\%, respectively, across all three lines), while AGN stacks are rarely masked (0--7\%).
This correction is applied independently to [O~II], [O~III], and [Ne~III]~using the procedures described above, and the resulting corrected data is used when comparing to AGN photoionization models in \S\ref{sec:ner_or}.

Table~\ref{table:sfg_correction_strength} summarizes the typical strength of this correction across the Ne-O sample: the star-forming contribution is largest for [O~II], modest for [O~III], and intermediate for [Ne~III], consistent with [O~II]~being the line most readily produced by star formation.
Because \ORs\ and \NeRs\ each combine two independently corrected lines, we also report the resulting shift in each ratio.

\begin{table}[ht]
\centering
\small
\caption{Star-forming contamination correction strength for the Ne-O sample. For each line, the fractional correction is $(F_{\rm uncorr} - F_{\rm corr})/F_{\rm uncorr}$; for the line-ratio rows, the shift is reported in dex ($\log_{10}(\text{corrected}) - \text{uncorrected}$). }
\label{table:sfg_correction_strength}
\begin{tabular}{||p{2.6cm} c p{1.6cm} p{1.8cm}||}
\hline
Quantity & N & Median Correction & Interquartile Range \\
\hline\hline
{[O~II]} flux & 2,006 & 13.5\% & 8.0\%--22.0\% \\
{[O~III]} flux & 2,034 & 1.8\% & 1.0\%--3.1\% \\
{[Ne~III]} flux & 2,036 & 2.3\% & 1.1\%--4.2\% \\
\hline
$\log$([O~III]/[O~II]) shift (dex) & 2,006 & 0.053 & 0.030--0.094 \\
$\log$([Ne~V]/[Ne~III]) shift (dex) & 2,036 & 0.010 & 0.005--0.019 \\
\hline
\end{tabular}
\end{table}

We do not, by contrast, attempt to correct \OII, \OIII, or \NeIII\ for dust reddening via the Balmer decrement.
A Balmer-decrement dust correction requires assuming an intrinsic $\mathrm{H}\alpha/\mathrm{H}\beta$ ratio, and our own AGN photoionization model grid (\S\ref{subsubsec:agn_models}) shows that the standard Case B value ($2.86$) does not hold for narrow-line-region gas. 
Models spanning the \NeRs--\ORs~range of our sample predict a mean intrinsic $\mathrm{H}\alpha/\mathrm{H}\beta = 3.06 \pm 0.17$, with individual models departing from this value in a manner strongly anti-correlated with black hole mass ($r = -0.81$).
To avoid introducing a systematic error that depends on black hole mass, we do not apply a dust correction to \NeR\ or \OR\ anywhere in this paper.  
Where relevant, we show reddening vectors (e.g., Figure~\ref{fig:ne53_o32_models}, Panel A), to illustrate the direction and approximate magnitude that dust would shift the data.

The sole exception to this is the $L_{\rm [OIII]} \to L_{\rm bol}$ conversion (and, by extension, \edd; \S\ref{sec:Data}), where we do apply a dust correction, computed from the Balmer decrement using a model-derived value of $\mathrm{H}\alpha/\mathrm{H}\beta = 3.06$ as the intrinsic value.
The case-B correction would boost NeV-AGN's median $L_{\rm [OIII]}$ by a factor of $4.1$ ($0.61$~dex), compared to a factor of $3.3$ ($0.51$~dex) using our updated value -- a difference of $0.10$~dex, or roughly 26\%, in the inferred luminosity.

%------------------------------------------------------------------------------------------------------
%                                                                                                    --
%                                          Section 4                                                 --
%                                                                                                    --
%------------------------------------------------------------------------------------------------------

\section{The BPT Sample: NeV-AGN and NC-AGN} \label{sec:nev-ncl}

\begin{table*}[ht]
\centering
\caption{[Ne~V] detection rate by BPT classification \citep{LawBPT}, for the BPT-classifiable sample of \Ha, \Hb, \NII, and \OIII~S/N $>$ 3, SPECPRIMARY =1, and $0.147 < z < 0.55$ (N = 60,521).}
\label{table:nev_by_bpt}
\begin{tabular}{||l c c c c||}
\hline
BPT Class & Number of Galaxies & Number with [Ne~V] & \% of Class with [Ne~V] & \% of [Ne~V]~in Class \\
\hline\hline
AGN            &  7,433 & 3,088 & 41.5\% &  83.1\% \\
Star-forming   & 26,442 &    32 &  0.1\% &   0.9\% \\
Composite      & 11,035 &   187 &  1.7\% &   5.0\% \\
LIER           & 15,611 &   411 &  2.6\% &  11.1\% \\
\hline
All four classes & 60,521 & 3,718 & 6.1\% & 100.0\% \\
\hline
\end{tabular}
\end{table*}

\begin{figure*}[!ht]
\centering
\includegraphics[width=0.9\textwidth]{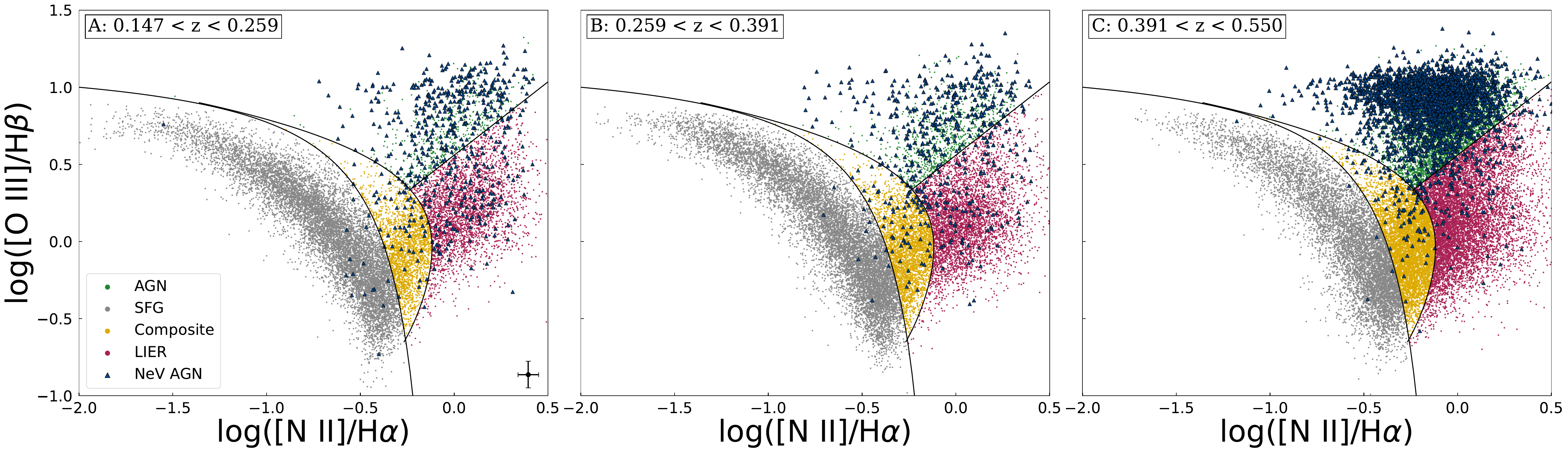}
\caption{The Baldwin, Phillips, and Terlevich (BPT) \citep{OG_BPT} diagnostic diagram of our sample split into three bins spanning equal intervals in the age of the universe ($0.147 < z < 0.259$, $0.259 < z < 0.391$, and $0.391 < z < 0.550$) for galaxies with H$\beta$, [O~III], H$\alpha$, and [N~II]~S/N $>$ 3.
The diagram is divided into regions following \citet{LawBPT}.
Galaxies with [Ne~V] detections of S/N $> 5$ are shown as navy triangles, Active Galactic Nuclei (AGN) are shown in green, Star-forming Galaxies (SFGs) are shown in gray, composite galaxies are shown in gold, and Low Ionization Emission Region (LIER) galaxies are shown in dark red. 
[Ne~V] detection statistics for this subsample can be found in Table~\ref{table:nev_by_bpt}.
\label{fig:bpt}}
\end{figure*}

\begin{figure}[ht]
\plotone{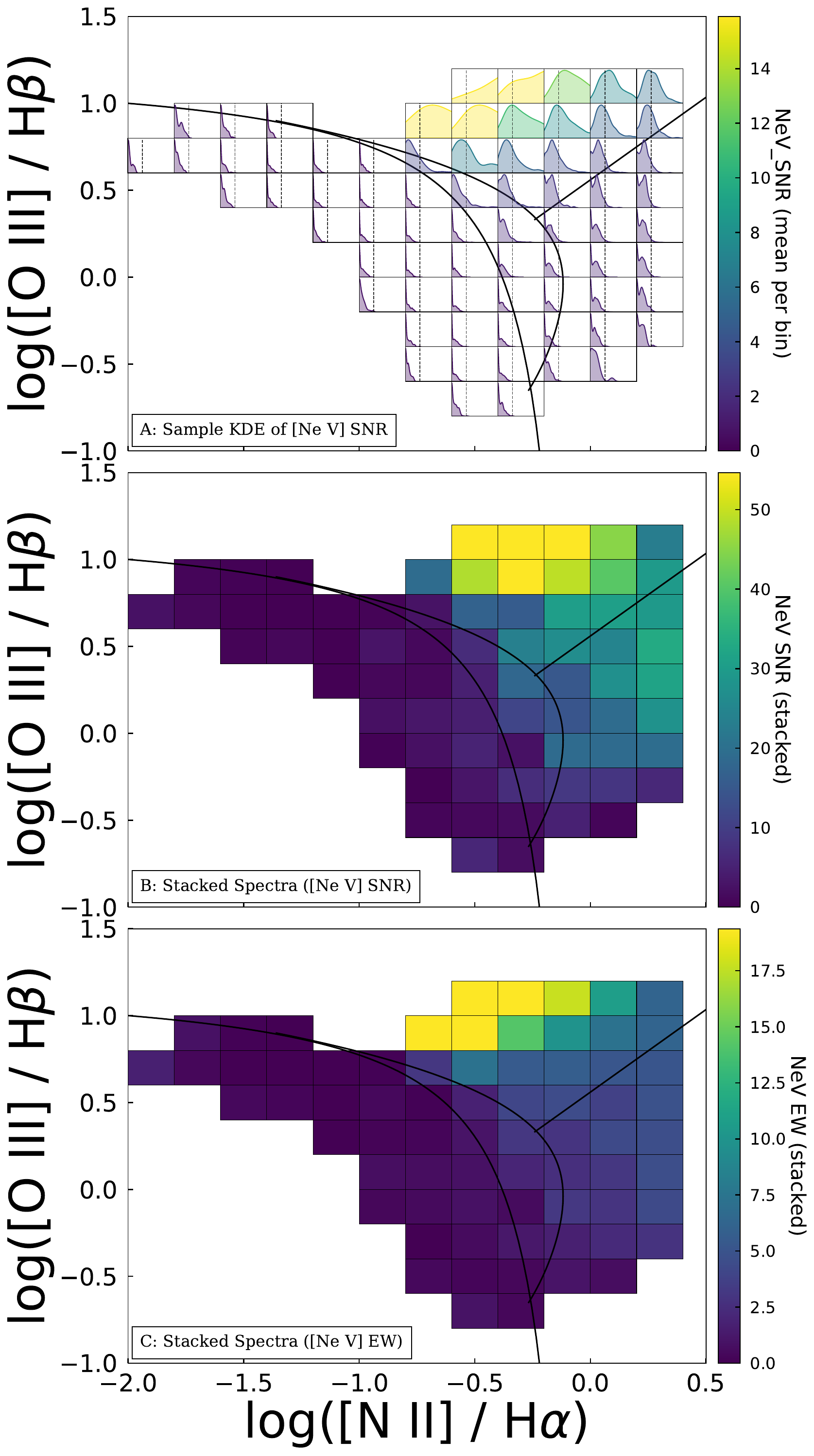}
\caption{BPT diagram
binned on a 0.2 dex grid in log ([N II] / H$\alpha$) and log ([O III] / H$\beta$); only bins with N$_{\mathrm{Galaxy}} \geq 20$ are shown.
\textbf{Panel A}: 1-D KDEs of the [Ne~V] S/N distribution in each bin, identical to Figure \ref{fig:mbh-loiii-agn-grid}. 
\textbf{Panel B}: The \NeVp~S/N in stacks made from between 20 and 500 galaxies per bin (see \S\ref{subsec:stacking_methodology}).
\textbf{Panel C}: The \NeVp~equivalent width in stacks made using the same binning and sampling as Panel B.
\label{fig:bpt_grid}}
\end{figure}

To characterize which AGN produce detectable \NeV~emission and which do not, we first examine the distribution of \NeVp~detections across the BPT diagram.
In Figure~\ref{fig:bpt}, we show the BPT diagram for our sample (N = 60,521) divided into three redshift bins spanning equal intervals in the age of the universe.
Galaxies with \NeVp~S/N $>$ 5 are shown in black, with the remaining BPT classifications following \citet{LawBPT}. Classification statistics are recorded in Table~\ref{table:nev_by_bpt}.
We find that 41.5\% of BPT AGN show detectable [Ne~V]~emission at S/N $>$ 5.
Of the 3,718 galaxies with [Ne~V]~S/N $>$ 5 across all BPT classes, 3,088 (83.1\%) are classified as BPT AGN, 598 (16.1\%) are composites or LIERs, and the remaining 32 (0.9\%) are star-forming galaxies.

In Figure~\ref{fig:bpt_grid}, we show the \NeV~S/N and equivalent width (EW) measured on stacked spectra spanning the BPT diagram.
Stacks were created as described in \S\ref{subsec:stacking_methodology} in bins of 0.2 dex in both \Ntwo~and \Rthree, using galaxies drawn from the sample described in \S\ref{subsec:nev-ncl_data} with $z < 0.55$.
Bins with fewer than 20 galaxies are excluded; bins with more than 500 are randomly subsampled to 500 without replacement.
The top panel shows 1-D KDEs of the \NeV~S/N distribution within each bin, with a vertical line at S/N $= 5$ marking the detection threshold adopted in this work. The middle and bottom panels show the \NeV~S/N and EW  measured on the stacked spectra, respectively.

Together, Figures \ref{fig:bpt} and \ref{fig:bpt_grid} establish that [Ne~V]\ emission is predominantly an AGN phenomenon, and that its strength tracks position within the BPT diagram.
Both stacked S/N and EW peak in the high [O~III]/H$\beta$, low [N~II]/H$\alpha$ corner of the diagram and decline smoothly toward the star-forming region.  Although [Ne\,V] is generally considered a weak emission line, its EW exceeds 15~\AA\ in several bins.  
The high [O\,III]/H$\beta$, low [N~II]/H$\alpha$ region of the BPT where the [Ne\,V] EW peaks is associated with AGN exhibiting high ionization parameters and/or low metallicities \citep{Thomas2019}, and is more commonly populated at higher redshift \citep{Kewley2013}.
Stacking reveals detectable [Ne\,V]\ extending into the composite and LIER regions, where individual spectra alone rarely reach our detection threshold. 
Lower-luminosity AGN populate these regions of the BPT diagram, and they show correspondingly weaker [Ne\,V]~emission (median EW 3.81~\AA\ and 5.47~\AA, respectively).

%------------------------------------------------------------------------------------------------------
%                                                                                                    --
%                                          Section 5                                                 --
%                                                                                                    --
%------------------------------------------------------------------------------------------------------

\section{Population Comparison} \label{sec:pop_comp}

The BPT diagram establishes that a large fraction of eBOSS AGN lack detectable [Ne~V] emission, but does not reveal why.
To explore this question, we defined samples of [Ne~V]-detected and undetected AGN (the NeV-AGN and NC-AGN) in \S\ref{subsec:nev-ncl_data}.
We compare the two populations across host galaxy properties (\S\ref{subsec:host}) and AGN properties (\S\ref{subsec:agn}) to identify whether the NeV-AGN/NC-AGN dichotomy reflects AGN conditions, host galaxy characteristics, or both.
We then stack NC-AGN spectra to test directly whether coronal emission is present in this population at levels too faint to detect individually (\S\ref{subsec:ncl-stacking}).

%---------------------------------------------------------------------------------------------------
%                                  Subsection 5.1                                                 --
%---------------------------------------------------------------------------------------------------

\subsection{Host Galaxy Properties} \label{subsec:host}

\begin{figure*}[ht]
\plotone{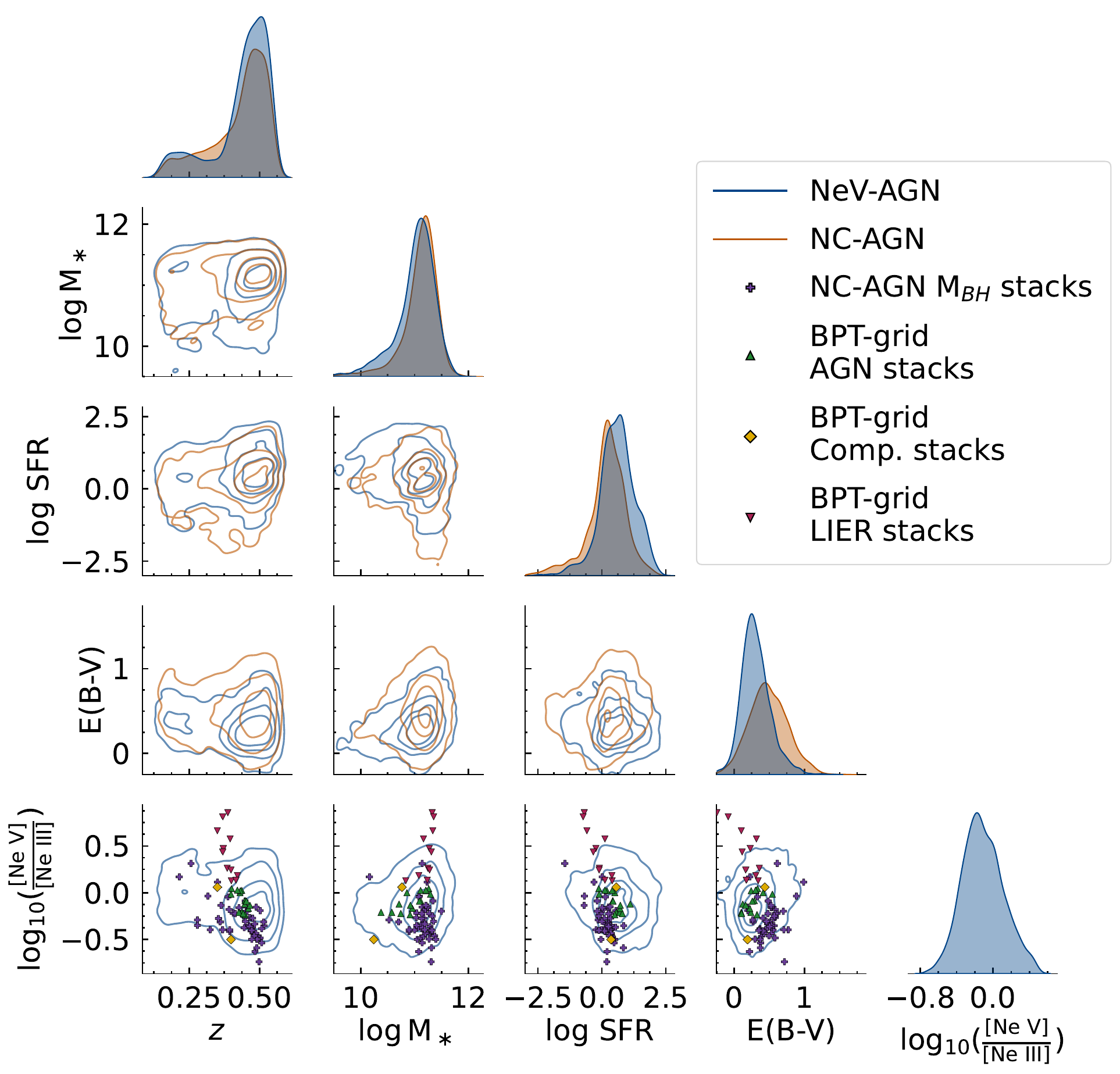}
\caption{Comparison of host galaxy properties for NeV-AGN (blue contours; N = 3,088) and Non-Coronal AGN (NC-AGN) or AGN that have no significant [Ne V] detection (orange contours; N = 2,626).
The corner plot shows the joint distributions of redshift, log(M$_{*}$), log SFR, E(B-V), and \NeR with kernel density estimate contours and diagonal histograms.
NC-AGN stacks (\S\ref{subsec:ncl-stacking}, purple crosses), BPT-grid AGN stacks (green triangles), BPT-grid composite stacks (yellow diamonds), and BPT-grid LIER stacks (maroon inverted triangles) are overlaid on the \NeR\ panels.
These panels are shown for NeV-AGN only, since NC-AGN by definition have [Ne~V]~S/N $<$ 3.
Stellar masses and star formation rates are from \drake~using CIGALE \citep{CIGALE} SED fitting; E(B-V), by contrast, is measured directly from the Balmer decrement (\S\ref{sec:Data}).
The two populations differ most in star formation rate and dust content, with NC-AGN systematically dustier and less actively star-forming than NeV-AGN (Table~\ref{table:ks_pop}).
\label{fig:nev-ncl-host-corner}}
\end{figure*}

We first ask whether NeV-AGN and NC-AGN occupy distinct regions of host galaxy parameter space.
Figure~\ref{fig:nev-ncl-host-corner} shows the joint distributions of redshift, log(M$_{*}$), log SFR, E(B-V), and \NeR\ (not corrected for star formation, unlike the classification described in \S\ref{sec:ner_or}) for the two populations.
The [Ne~V]/[Ne~III] panels are shown for NeV-AGN only, since NC-AGN by definition have \NeV~S/N $<$ 3. NC-AGN stacks binned by black hole mass and [O~III] luminosity (\S\ref{subsec:ncl-stacking}), and BPT-grid stacks (\S\ref{sec:nev-ncl}) separated into AGN, composite, and LIER classes, are overlaid on these panels.
Stellar masses and star formation rates are measured by \drake~using CIGALE \citep{CIGALE} SED fitting as described in \S\ref{sec:Data}.
E(B-V), by contrast, is measured directly from the Balmer decrement, $E(B-V) = 1.97 \log_{10}[(\mathrm{H}\alpha/\mathrm{H}\beta)/3.06]$, using the \citet{Calzetti} attenuation law and the AGN-model-derived intrinsic ratio described in \S\ref{sec:methods_extra}, rather than the standard case-B value.

We find that the redshift distributions of NeV-AGN and NC-AGN peak at nearly the same value (0.506 vs. 0.484, a 4.4\% difference; see Table~\ref{table:ks_pop}), while their stellar mass distributions differ more substantially: NC-AGN peak at a stellar mass approximately 21.3\% higher than NeV-AGN ($\log M_{*} = 11.206$ vs.\ $11.121$, a separation of 0.084~dex; Table~\ref{table:ks_pop}).
The two populations differ most strongly in star formation rate and dust content: NeV-AGN peak at a star formation rate approximately 3.1 times (207\%) higher than NC-AGN at fixed stellar mass ($\log \mathrm{SFR} = 0.710$ vs.\ $0.223$, a separation of 0.487~dex; Table~\ref{table:ks_pop}), and NC-AGN are significantly dustier, peaking at an E(B-V) approximately 55.4\% higher than NeV-AGN ($0.440$ vs.\ $0.249$~mag, a separation of 0.191~mag; Table~\ref{table:ks_pop}).
Curiously, \NeR\ shows a weak \emph{positive} trend with E(B-V) (slope $0.192$, 95\% CI $[0.149, 0.235]$), but E(B-V) accounts for only 3\% of the variance in $\log$([Ne\,V]/[Ne\,III]) ($R^2 = 0.030$), leaving most of the scatter unexplained by dust alone.
Notably, several NC-AGN stacks extend into regions of the [Ne\,V]/[Ne\,III] parameter space not populated by individual NeV-AGN, a trend we examine in more detail in \S\ref{subsec:ncl-stacking}.
Full KS statistics for all host galaxy properties are reported in Appendix~\ref{apx:ks}.

Taken together, these results paint a complicated picture of dust's role in coronal line emission: NC-AGN are markedly dustier than NeV-AGN at the population level (E(B-V) is higher by 55.4\%; see above), yet dust content explains almost none of the [Ne\,V]/[Ne\,III] scatter among NeV-AGN.  Indeed, the weak positive trend between [Ne\,V]/[Ne\,III] and E(B-V) has the opposite sign to that expected from a simple foreground dust screen.
One possible explanation lies in the geometry of the coronal line region (CLR), which sits interior to the more extended [Ne\,III]-emitting region \citep{clr_location}.
If dust preferentially attenuates the more extended, lower-ionization gas, the net effect could be an apparent strengthening of \NeR\ with increasing dust content, the reverse of the naive expectation.
However, it is more likely that our results reflect two competing selection effects rather than this geometric mechanism.
Within the NeV-AGN sample, the weak positive trend may arise because [Ne~V] is more readily detected in the most luminous AGN, which tend to reside in more strongly star-forming and dustier hosts.
At the population level, however, [Ne~V] becomes too weak to detect in the dustiest galaxies at fixed star formation rate (Fig.~\ref{fig:nev-ncl-host-corner}), so NC-AGN end up dustier on average than NeV-AGN.

%---------------------------------------------------------------------------------------------------
%                                  Subsection 5.2                                                 --
%---------------------------------------------------------------------------------------------------

\subsection{AGN Properties} \label{subsec:agn}

\begin{figure*}[ht]
\plotone{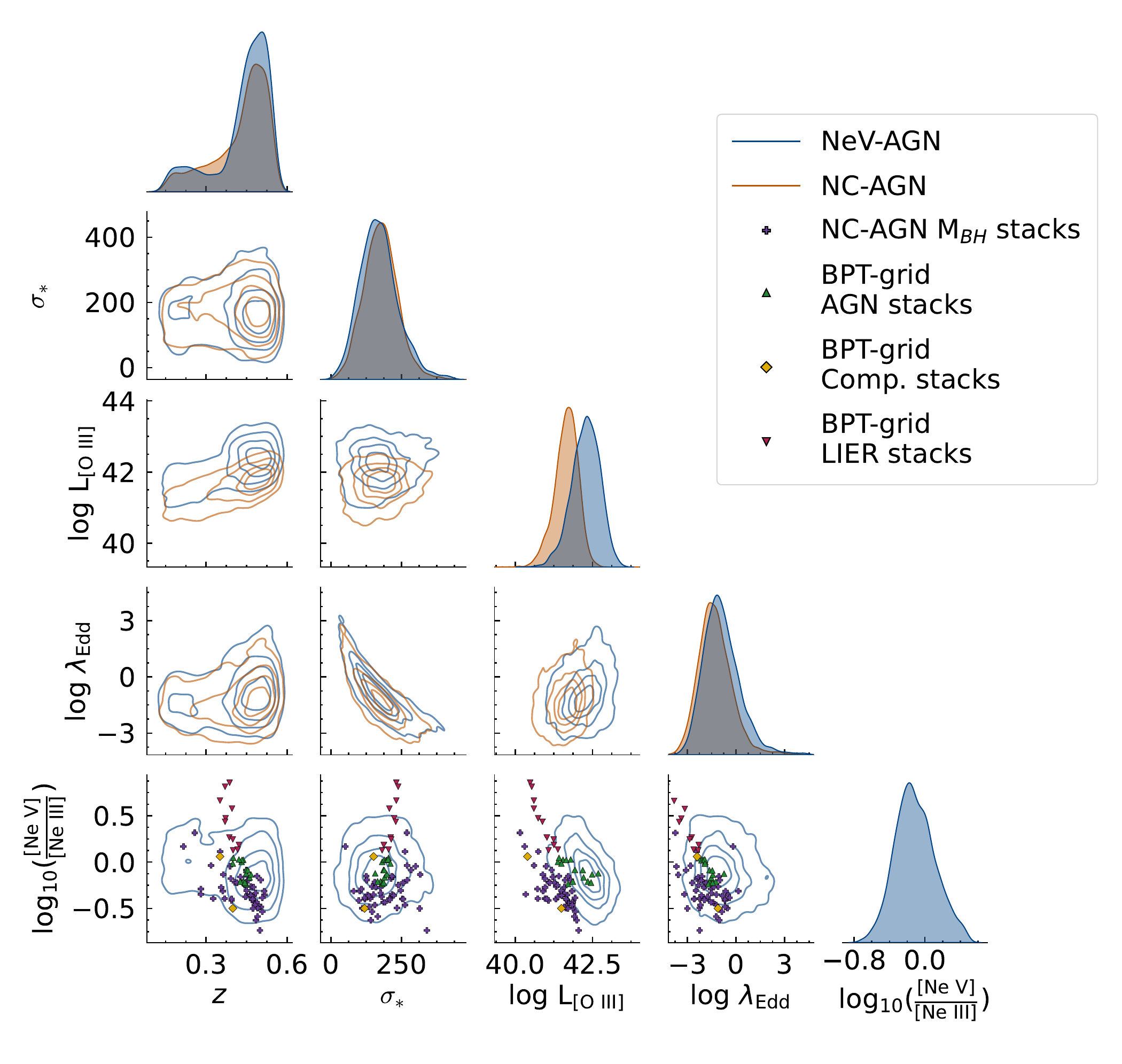}
\caption{Comparison of AGN properties for NeV-AGN (blue contours; N = 3,088) and NC-AGN (orange contours; N = 2,626).
The corner plot shows the joint distributions of redshift, stellar velocity dispersion ($\sigma_{*}$; a proxy for black hole mass), log \OIII~luminosity ($\log L_{\rm [OIII]}$; a proxy for the AGN accretion rate), log Eddington ratio(\Ledd), and \NeR\ (star-formation corrected; see \S\ref{subsec:sfg_matching}), with kernel density estimate contours and diagonal histograms.
The $\sigma_{*}$ and Eddington ratio panels are restricted to $\sigma_{*}$ uncertainty $< 50~\mathrm{km~s^{-1}}$.
NC-AGN stacks (\S\ref{subsec:ncl-stacking}, purple crosses), BPT-grid AGN stacks (green triangles), BPT-grid composite stacks (yellow diamonds), and BPT-grid LIER stacks (maroon inverted triangles) are overlaid on the \NeR\ panels.
The two populations differ most in [O\,III]~luminosity and Eddington ratio, with NeV-AGN more luminous and accreting faster than NC-AGN (Table~\ref{table:ks_pop}).
\label{fig:nev-ncl-agn-corner}}
\end{figure*}

We next compare the AGN properties of the two populations, as shown in Figure~\ref{fig:nev-ncl-agn-corner}: redshift, stellar velocity dispersion ($\sigma_{*}$), \OIII~luminosity ($L_{[O~III]}$), Eddington ratio(\edd), and \NeR.
The velocity dispersion serves as a proxy for black hole mass via the M$_{\rm BH}$--$\sigma_{*}$ relation (\S\ref{sec:Data}), allowing us to test whether coronal line detectability depends on the mass of the central black hole.
$L_{\rm [OIII]}$ serves as a proxy for AGN bolometric luminosity via a fixed correction factor and is among the most reliable bolometric indicators for Type II AGN \citep{Lbol}.
Eddington ratio combines both quantities to test whether detectability instead depends on accretion rate relative to the Eddington limit.

We find that NeV-AGN and NC-AGN differ most strongly in \OIII~luminosity and Eddington ratio.
NeV-AGN peak at an [O~III] luminosity approximately 3.9 times (294\%) higher than NC-AGN ($\log L_{\rm [O\,III]} = 42.327$ vs.\ $41.731$, a separation of 0.595~dex).
NC-AGN stacks (\S\ref{subsec:ncl-stacking}) and BPT-grid stacks extend the observed $L_{\rm [OIII]}$--(\NeRs) relation to lower luminosities than any individual NeV-AGN, with LIER stacks reaching $\log$~(\NeRs) values systematically above the full NeV-AGN range (median offset $+0.677$~dex).

NeV-AGN accrete at Eddington ratios approximately 2.8 times (175\%) higher than NC-AGN (\Ledd$ = -1.164$ vs.\ $-1.604$, a separation of 0.440~dex); since Eddington ratio at fixed black hole mass tracks bolometric luminosity, this is consistent with NeV-AGN simply producing a larger total budget of $>$97~eV ionizing photons, sufficient to cross our detection threshold, independent of the shape of the ionizing continuum -- we return to this distinction, and to the shape of the continuum itself, in \S\ref{sec:ner_or}.
The NC-AGN and NeV-AGN nonetheless overlap substantially in $\log \lambda_{Edd}$ (medians of $-1.010$ and $-1.394$, with standard deviations of order 1.1~dex in each population). Hence, this separation reflects a real but modest shift in typical accretion rate rather than a clean division between the populations.
The stellar velocity dispersion, $\sigma_{*}$, differs by a comparatively modest 13.3\% (156.1 vs.\ 178.4~\kms, a separation of 22.3~\kms), suggesting that the black hole masses are similar in the two samples.
Full KS statistics for all AGN properties are reported in Appendix~\ref{apx:ks}.

In summary, NeV-AGN are both more luminous (approximately 3.9$\times$ in $L_{\rm [OIII]}$) and accrete at higher Eddington ratios (approximately 2.8$\times$) than NC-AGN, while having a similar distribution of black hole masses.

%---------------------------------------------------------------------------------------------------
%                                  Subsection 5.3                                                 --
%---------------------------------------------------------------------------------------------------

\subsection{NC-AGN Stacking} \label{subsec:ncl-stacking}

The [Ne~V] non-detections that define NC-AGN could reflect either a genuine physical absence of coronal emission, or simply a sensitivity limitation: individual NC-AGN spectra may host coronal emission too faint to detect at our S/N threshold.
We test this directly by constructing stacks identical to the AGN stacks shown in Figure~\ref{fig:mbh-loiii-agn-grid} (\S\ref{subsec:stacking_methodology}) -- the same $M_{\rm BH} \times L_{\rm [OIII]}$ binning and $\sigma_{*}$ quality cuts -- but restricted to NC-AGN only (\NeVp~S/N $<$ 3).

Stacking recovers a formal [Ne~V] detection (S/N $> 5$) in 56 of the 83 bins with at least five contributing galaxies (67.5\%), rising to 31 of 37 bins (83.8\%) once bins with fewer than 20 galaxies are excluded -- despite every individual parent spectrum falling below our S/N $<$ 3 threshold by construction. Detections concentrate at intermediate to high black hole mass ($\log M_{\rm BH} \sim 7$--$9$) and higher \OIII~luminosity ($\log L_{\rm [OIII]} \gtrsim 41.5$); the lowest-mass bins are also our most sparsely populated, consistent with the remaining non-detections reflecting a sensitivity limitation rather than a physical cutoff in coronal-line production. Equivalent width follows a broadly similar pattern to S/N.

The stacks are offset from the individual populations in Figures~\ref{fig:nev-ncl-host-corner} and \ref{fig:nev-ncl-agn-corner}, fainter in \OIII~luminosity ($0.65$~dex below the combined individual sample), with a dust content (E(B-V) $= 0.490$) comparable to individual NC-AGN ($0.440$) and higher than individual NeV-AGN ($0.249$).

Taken together, these results indicate that the NeV-AGN/NC-AGN dichotomy is primarily a sensitivity effect rather than a fundamental difference in the conditions required to produce coronal line emission.
Having established that \OIII~luminosity and Eddington ratio, rather than black hole mass, are most closely tied to coronal line detectability, we next examine the ionization structure of the sample directly using [Ne~V]/[Ne~III] and [O~III]/[O~II] line ratios in \S\ref{sec:ner_or}.

%------------------------------------------------------------------------------------------------------
%                                                                                                    --
%                                          Section 6                                                 --
%                                                                                                    --
%------------------------------------------------------------------------------------------------------

\section{The [Ne V] / [Ne III] -- [O III] / [O II] Diagnostic} \label{sec:ner_or}

We now turn to better understanding the shape of the AGN SED by exploring the ratio of [Ne~V] to other lines.
We use four emission lines that probe the shape of the ionizing continuum from roughly 15 to 100~eV: \OII~(13.62 eV), \OIII~(35.12 eV), \NeIII~(40.96 eV), and [Ne~V]~3427 (97.1 eV).

\begin{figure*}
\centering
\includegraphics[width=\textwidth]{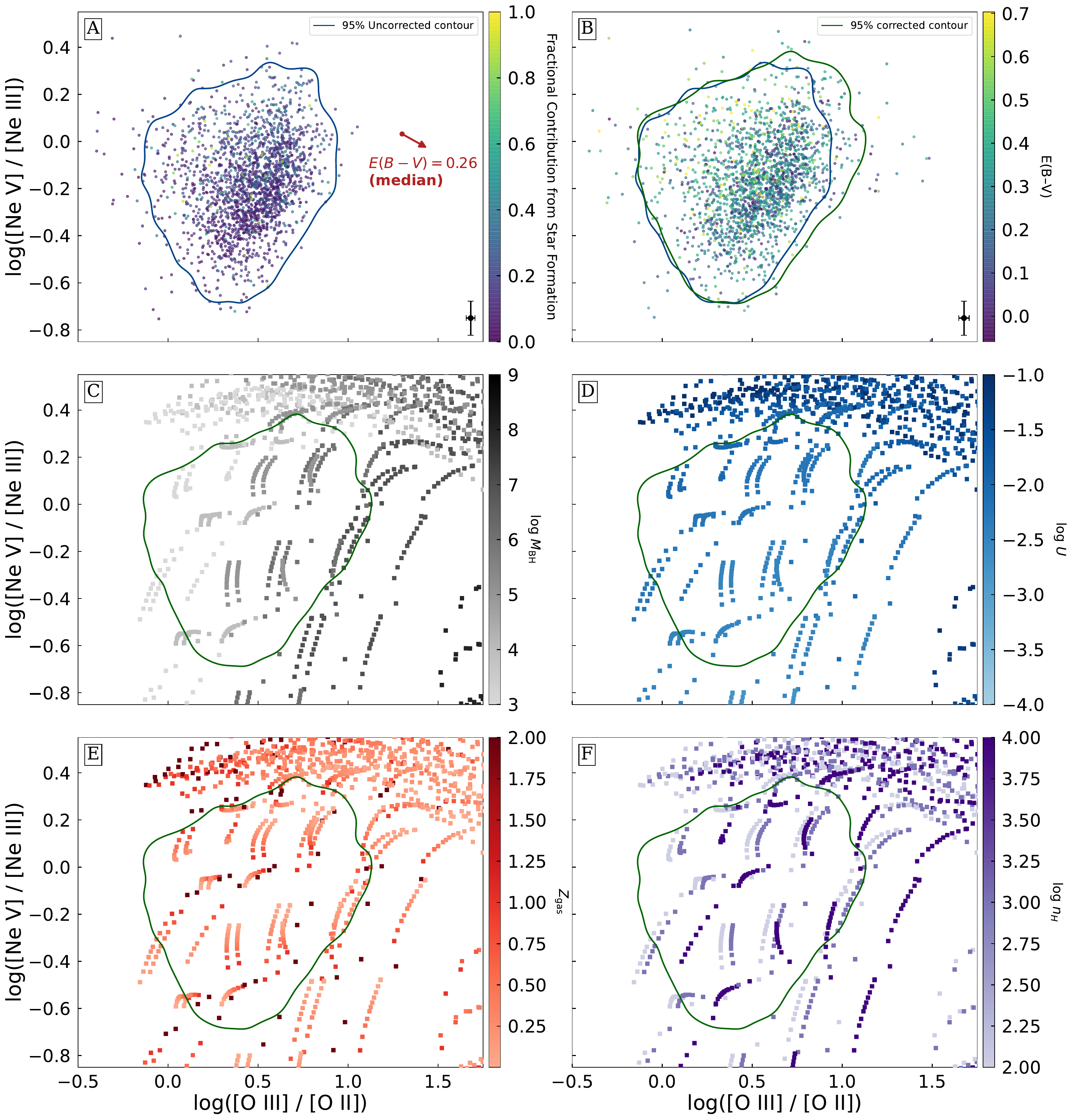}
\caption{\NeR\ vs.\ \OR\ for the Ne-O sample (see \ref{subsec:ne-o_data}), compared with AGN photoionization models from \citet{cleri2025}.
The blue (Panels A, B) and dark green contours (Panels B--F) enclose 95\% of the data uncorrected and corrected for star formation, respectively (see \S\ref{subsec:sfg_matching}).
\textbf{Panel A:} uncorrected line ratios of the Ne-O sample, color-coded by the fractional star-forming contribution to [O~II]. The red arrow shows a foreground reddening vector of $E(B-V) = 0.27$, the sample median.
\textbf{Panel B:} The same sample as Panel A, but now the emission line ratios have been corrected for star formation contamination. Individual points are color-coded by E(B-V), while the contours summarize the population distributions of the uncorrected (blue) and corrected (dark green) samples.
\textbf{Panels C--F:} AGN photoionization models from \citet[][see \S\ref{subsubsec:agn_models}]{cleri2025} color-coded by (C) black hole mass, (D) ionization parameter, log\,U, (E) gas-phase metallicity, $Z_{\rm gas}$, and (F) hydrogen density, $\log(n_{\rm H}/\mathrm{cm}^{-3})$.
The 95\% contour of the corrected data is shown in green for comparison.
The dominant variation in line ratios across Panels C--F arises from the combination of black hole mass and ionization parameter, log\,U.
\label{fig:ne53_o32_models}}
\end{figure*}

The Ne-O sample comprises 2,458 galaxies meeting S/N $>$ 5 on \NeVp, \OIIp, \OIII, and \NeIII, drawn from the redshift range $0.147 < z < 0.55$.
[O\,II], [O\,III], and [Ne\,III] are each susceptible to star-formation contamination, so we correct all three using the SFG matching and stacking procedure described in \S\ref{sec:methods_extra}.
This correction succeeds for 2,196 of 2,458 galaxies (89.3\%); the remainder either lack sufficient SFG matches or have an unusable corrected value for at least one line, and are excluded from all corrected quantities.
The star-forming contribution is largest for \OII~(median 13.5\%), modest for \OIII~(median 1.8\%), and intermediate for \NeIII~(median 2.3\%).
[Ne\,V] itself requires no correction given the very small amount of [Ne\,V] produced by stars (see Fig.~\ref{fig:ne53_o32_full_grid}).
We show the distribution of the Ne-O sample in the \NeR--\OR\ plane in panel A of Figure~\ref{fig:ne53_o32_models}; line ratios corrected for star formation contamination are shown in panel B.

A naive expectation is that [O\,III]/[O\,II]~and [Ne\,V]/[Ne\,III] should correlate tightly, since both trace the hardness of the ionizing continuum, but in different energy regimes; instead, the corrected Ne-O sample scatters broadly, indicating that AGN in our sample span a genuinely diverse range of high-ionization continuum shapes.
Figure~\ref{fig:ne53_o32_models} (Panels C--F) shows that AGN photoionization model predictions from
\citet{cleri2025} span roughly the same parameter space when considering variations in black hole mass, ionization parameter, gas-phase metallicity, and hydrogen column density (see \S\ref{subsubsec:agn_models} for model details). No single parameter cleanly separates the observed scatter along either axis, though black hole mass and ionization parameter together account for most of the variation (Figure~\ref{fig:ne53_o32_models}).

%---------------------------------------------------------------------------------------------------
%                                  Subsection 6.1                                                 --
%---------------------------------------------------------------------------------------------------

\subsection{The High53/Locus/Low53 Classification} \label{subsec:hll}

\begin{figure}[ht]
\plotone{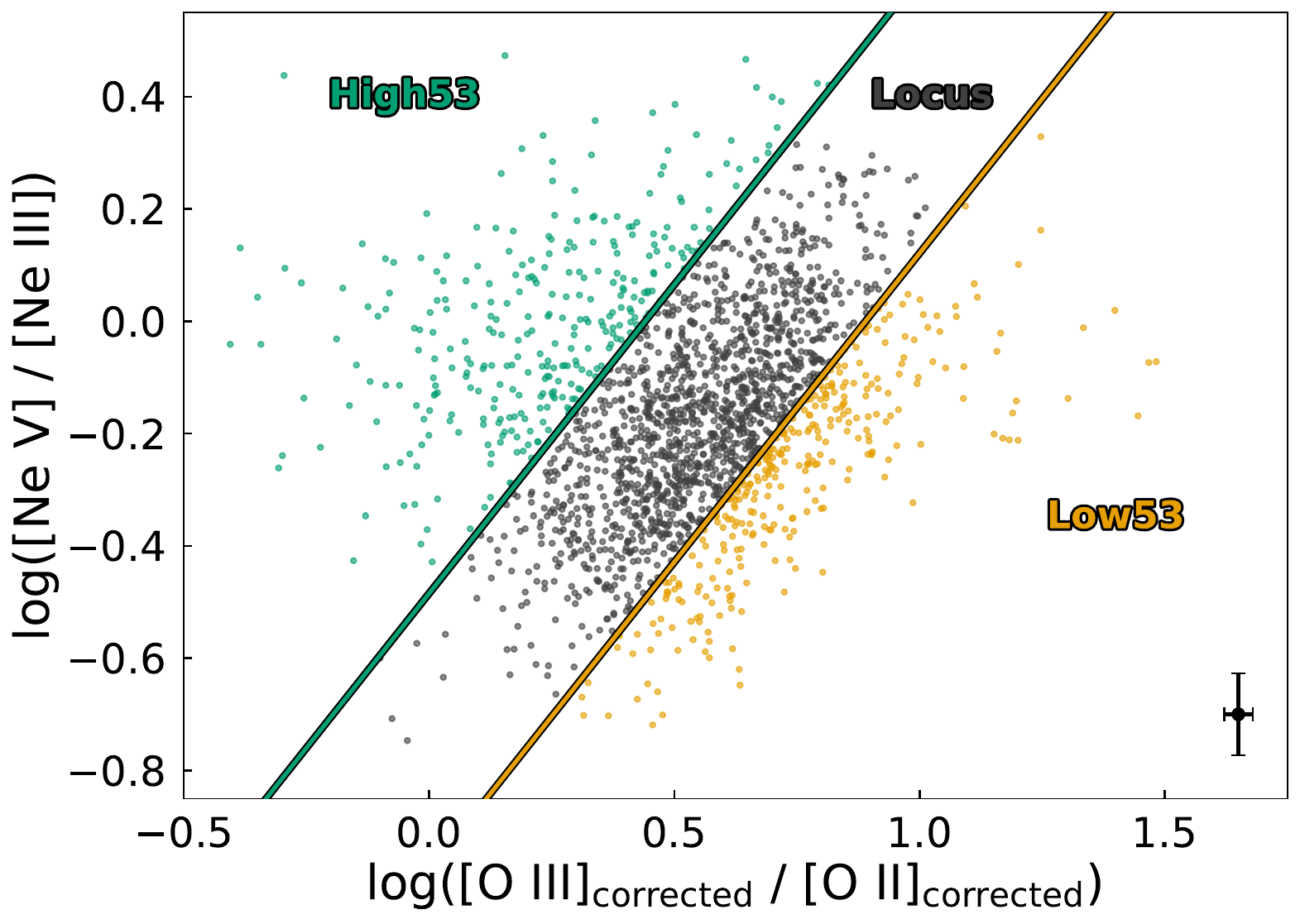}
\caption{\NeRcorrs~vs \ORcorrs~for galaxies in the Ne-O sample, after the star-forming correction to [O~II], [O~III], and [Ne~III]~described in \S\ref{subsec:sfg_matching}.
The data have been divided into three groups as described in \S\ref{subsec:hll}: High53 (green, N = 316), Locus (gray, N = 1,367), and Low53 (orange, N = 315).
Rather than a single tight sequence, the data show substantial scatter well in excess of the representative uncertainty point (black point with error bars), motivating a division into three groups whose physical properties and BPT classifications we compare in \S\ref{subsec:hll}.
\label{fig:Ne53-O32-selection}}
\end{figure}

Galaxies scatter significantly above and below the central locus of the \NeRcorrs --\ORcorrs~sequence.
We examine whether this scatter reflects distinct physical populations by dividing the sample into three groups, ``High53'', ``Locus'', and ``Low53'', as shown in Figure~\ref{fig:Ne53-O32-selection}, and comparing their BPT classifications and physical properties across the groups.

We chose a slope of 1.1 to approximate the overall slope of the \NeRcorrs--\ORcorrs~relation.
The intercepts were chosen so that the High53 and Low53 samples each contain 16\% of the data, yielding the following demarcation lines:
\begin{equation}
\log([\mathrm{Ne\,V}]/[\mathrm{Ne\,III}]_{\rm corr}) = 1.1 \times \log([\mathrm{O\,III}]_{\rm corr}/[\mathrm{O\,II}]_{\rm corr}) - 0.386
\label{eq:high_locus}
\end{equation}
\begin{equation}
\log([\mathrm{Ne\,V}]/[\mathrm{Ne\,III}]_{\rm corr}) = 1.1 \times \log([\mathrm{O\,III}]_{\rm corr}/[\mathrm{O\,II}]_{\rm corr}) - 0.886
\label{eq:locus_low}
\end{equation}
Equation~\ref{eq:high_locus} separates the ``High53'' and ``Locus'' groups, and Equation~\ref{eq:locus_low} separates the ``Locus'' and ``Low53'' groups.

\begin{figure*}[ht]
\plotone{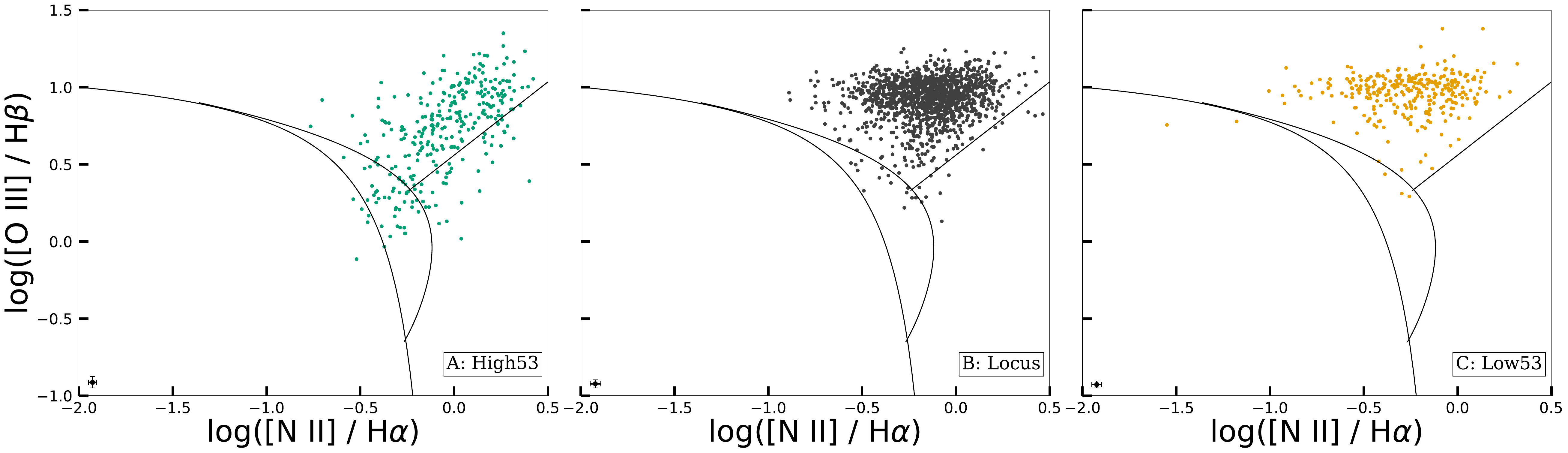}
\caption{BPT diagrams for the Ne-O galaxy subsample, divided into the ``High53'' (A, green), ``Locus'' (B, gray), and ``Low53'' (C, orange) groups defined by their position in the log ([Ne V] / [Ne III])~vs log ([O III]$_{\mathrm{corr}}$ / [O II]$_{\mathrm{corr}}$)~plane (Figure~\ref{fig:Ne53-O32-selection}; Equations~\ref{eq:high_locus}--\ref{eq:locus_low}).
Diagram regions follow \citet{LawBPT}.
97.4\% of the ``Locus'' group and 98.1\% of the ``Low53'' group fall within the BPT AGN region, compared to 72.2\% of the ``High53'' group.
\label{fig:O-Ne-bpt}}
\end{figure*}

Figure~\ref{fig:O-Ne-bpt} shows the BPT classification of the three groups.
We find that the Locus and Low53 groups are predominantly classified as BPT AGN (97.4\% and 98.1\%, respectively).
The High53 group, by contrast, is far more likely to contain LIERs and composite galaxies, with only 72.2\% falling within the BPT AGN region despite strong [Ne~V]~detections.
Excluding the High53 group therefore offers a way to select a comparatively pure AGN sample when the redder BPT lines, H$\alpha$ and [N~II], are unavailable.

\begin{figure*}[ht]
\plotone{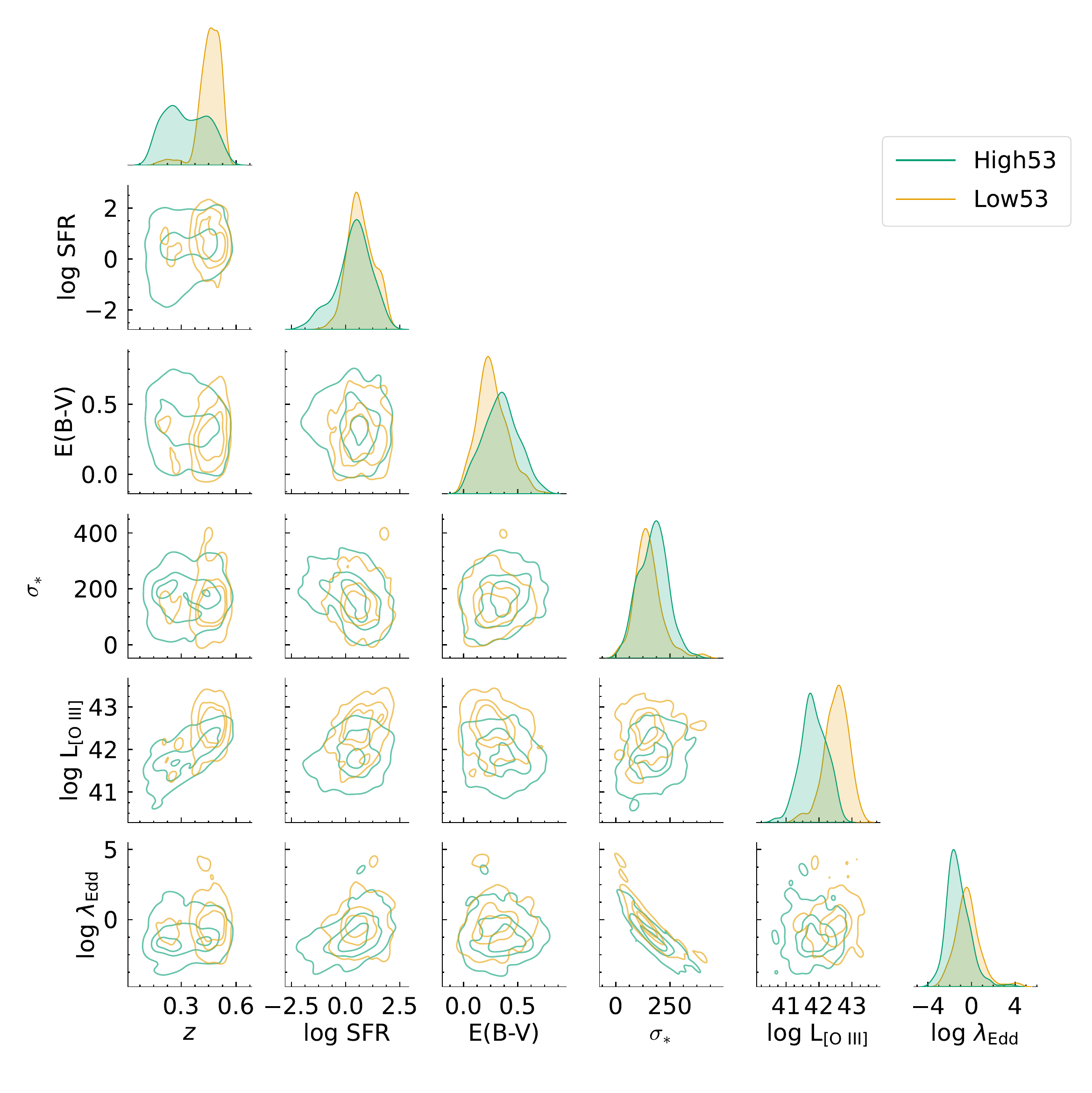}
\caption{Comparison of host galaxy and AGN properties for the ``High53'' (green, N = 316) and ``Low53'' (orange, N = 315) groups.
The corner plot shows the joint distributions of redshift, log SFR, E(B-V), $\sigma_{*}$, log $L_{\rm [OIII]}$, and \Ledd, with kernel density estimate contours and diagonal histograms.
The ``Locus'' group (N = 1,367) is omitted from this figure for visual clarity, but is included in the text discussion below and the full KS statistics reported in Appendix~\ref{apx:ks}.
The key trend is that Eddington ratio cleanly separates all three groups, decreasing from Low53 to Locus to High53, while [O~III] luminosity primarily distinguishes High53 from the other two.
\label{fig:hll-corner}}
\end{figure*}

In Figure~\ref{fig:hll-corner}, we compare the physical properties of the High53 and Low53 groups; KS statistics for all three groups can be found in Table~\ref{table:ks_neo}. We see that the redshift distributions of the Low53 and High53 groups differ dramatically: the Low53 group peaks sharply near $z \approx 0.46$, while the High53 group has an additional peak at $z \approx 0.25$.
The High53 group has higher E(B-V) ($0.350$ vs.\ $0.228$~mag) but relatively similar star formation rates.  Locus and Low53 are statistically indistinguishable from each other in both SFR and E(B-V) ($p = 0.29$ and $p = 0.13$).

We find that Eddington ratio cleanly orders all three groups: Low53 peaks highest (\Ledd$ = -0.375$), Locus intermediate (\Ledd$=-1.074$), and High53 lowest (\Ledd$=-1.650$), with every pairwise comparison statistically significant.
\OIII~luminosity, in contrast, separates High53 from the other two groups: it peaks $0.73$ and $0.87$~dex lower than Locus and Low53, respectively; Locus and Low53 differ from each other by only $0.13$~dex, a much smaller separation that is nonetheless formally significant given the large sample sizes ($p = 1.8\times10^{-4}$).
The stellar velocity dispersion, $\sigma_{*}$, a proxy for black hole mass, is highest for High53 ($187$~\kms) and lowest for Low53 ($137$~\kms).
These qualitative distinctions between groups persist under small variations of the slope and intercept of the demarcation line placement.

The extreme [Ne~V]/[Ne~III] values of the High53 group, despite its low [O~III]~luminosity, might naively suggest a small black hole mass, since lower-mass AGN in our model grid also occupy the high-Ne53 region of Figure~\ref{fig:ne53_o32_full_grid}; however, we find this is instead consistent with a low-Eddington-ratio, radiatively inefficient accretion state.
\citet{Kewley2006} find that LINERs occupy systematically lower Eddington ratios than Seyferts and require a harder ionizing radiation field, analogous to the low/hard state of X-ray binaries.
This picture naturally accommodates the High53 group's elevated $\sigma_{*}$ relative to Locus and Low53, since the mechanism depends on accretion rate rather than black hole mass. It is further supported by the High53 group's large fraction of LIER- and Composite-classifications (\S\ref{subsec:hll}).

The NC-AGN most closely resemble the High53 group in terms of their host and AGN physical properties.
To determine the [Ne~V] flux we would expect them to have, based on their detected [Ne~III], [O~III], and [O~II], we use Equation~\ref{eq:high_locus}, the demarcation line for the High53 group.  We then combine this flux value with the measured noise to determine the expected [Ne~V] S/N.   The median value for the NC-AGN sample is S/N=2.7, a non-detection.
However, 28.7\% (698 of 2,432) of NC-AGN have a predicted S/N $>5$ -- meaning that under our fairly conservative assumptions, they should have been detected but were not.  This may be due to additional dust reddening or a lower intrinsic [Ne~V]/[Ne~III] in this subset than assumed.

The physical origin of the Low53 group is less certain than that of the High53 group.
The Low53 group has the highest Eddington ratio of the three groups and significantly lower $\sigma_{*}$ than both Locus and High53, a combination consistent with a comparatively small black hole accreting rapidly.
This is in tension, however, with the AGN photoionization models occupying the Low53 group's region of the diagnostic plane, which require a median black hole mass of $\log (M_{\rm BH}/M_{\odot}) = 6$, compared to $\sim5$ for Locus and $\sim4$ for High53 (Figure~\ref{fig:ne53_o32_models}, Panel C; all pairwise comparisons statistically significant).
However, the models assume a fixed value of the Eddington ratio (\Ledd$ = -1$), a parameter which can also affect the hardness of the spectrum.
The Low53 group shows the largest [O~II] star formation correction of the three groups (mean fractional correction 25.4\%, versus 15.4\% for High53 and 15.9\% for Locus), consistent with these hosts having both high SFRs and a high black hole accretion rate.

We interpret the suppressed [Ne~V]/[Ne~III] in the Low53 group as reflecting a difference in the shape of the ionizing continuum itself.
The High53 group properties sit toward the low/hard end of this continuum, where hard-continuum, low-\edd\ nuclei produce disproportionately strong high-ionization emission.
The Low53 group properties sit toward the high/soft end, where disc-dominated, high-\edd\ nuclei suppress high-ionization emission \citep{highsoftlowhard, done2012}.
Unlike the often rapid, hysteretic state transitions seen in individual X-ray binaries, our sample shows a smooth, unimodal distribution in Eddington ratio, consistent with these three groups populating a continuum of accretion states rather than falling into two discrete regimes.
We caution that these interpretations are qualitative; a systematic comparison of the three group distributions against photoionization models across the complete accretion parameter space would be needed to place them on a quantitative footing, and we defer this comparison to future work.

In summary, [Ne~V]/[Ne~III] is not tightly correlated with [OIII]/[OII] (Fig.~\ref{fig:Ne53-O32-selection}). The scatter above and below the central locus reflects distinct physical populations rather than measurement noise, with the High53 and Low53 groups tracing different accretion regimes. The [Ne~V]/[Ne~III] vs. [OIII]/[OII] diagnostic is abundance-independent and can be used to identify populations with a higher fraction of composites and LIERs.

%------------------------------------------------------------------------------------------------------
%                                                                                                    --
%                                          Section 7                                                 --
%                                                                                                    --
%------------------------------------------------------------------------------------------------------

\section{The [Ne~V]/[O~III]--\texorpdfstring{$L_{\mathrm{[OIII]}}$}{L[OIII]} Anti-correlation} \label{sec:nev_oiii_loiii}

\begin{figure*}[ht]
\plotone{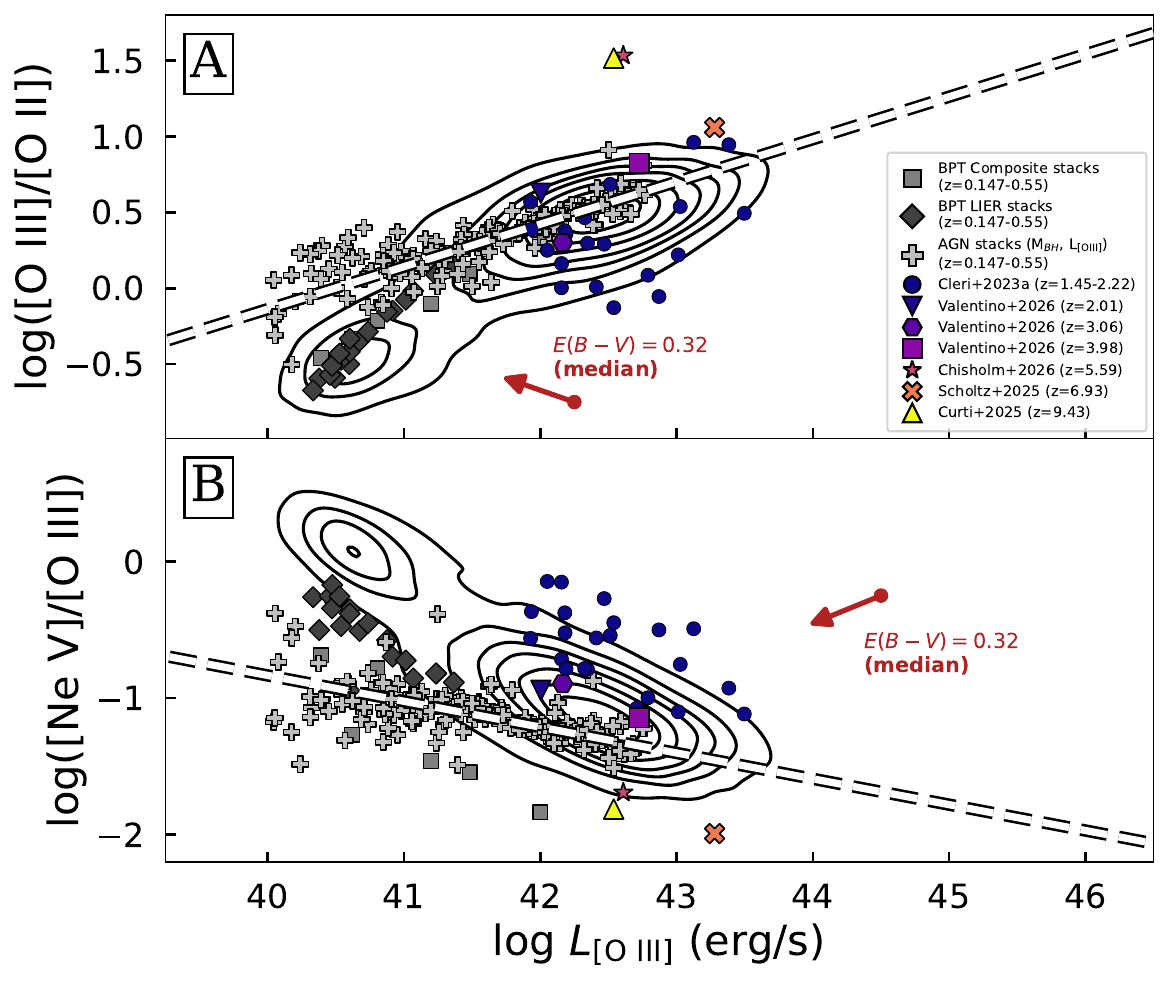}
\caption{Emission-line ratios versus AGN luminosity for individual eBOSS galaxies (black contours), eBOSS spectral stacks (grey symbols), and $z=1$--$9$ AGN from the literature (colored points).   Measurements made from AGN stacked on black hole mass and [O\,III] luminosity are shown as grey plus symbols, while measurements made from BPT Composite and LIER stacks are shown as grey squares and diamonds, respectively. The high redshift data includes 25 HST GRISM observations of $z=1.45-2.2$ AGN from the CLEAR sample \citep{CLeri2023} and 6 AGN observed with JWST whose properties are summarized in  
Table~\ref{table:nev_highz}.
\textbf{Top:} $\OR$ vs.\ $L_{\rm [OIII]}$ for the full [Ne~V]~parent sample with combined \NeVp~S/N $>$ 5 and no upper redshift limit, additionally requiring \OIII~and \OII~S/N $>$ 3 (N = 8,885). A weighted fit to the stacked spectra  (Equation~\ref{eq:o32_loiii_fit}) is shown as a dashed line.
\textbf{Bottom:} $\log([\mathrm{Ne\,V}]/[\mathrm{O\,III}])$ vs.\ $L_{\rm [OIII]}$ for the same sample and overlaid populations, with the corresponding fit to the stacked spectra (Equation~\ref{eq:nev_oiii_loiii_fit}) shown as a dashed line.
The reddening arrow shows the sample median $E(B-V) = 0.32$.  Data in this figure has not been corrected for reddening or star formation contamination.
\label{fig:o32_nev_oiii_loiii}}
\end{figure*}

The \NeRs--[O\,III]/[O\,III]~relation is a useful diagnostic of the shape of the AGN ionizing continuum, but it does not reveal how ionization state depends on AGN luminosity.
To test this directly, Figure~\ref{fig:o32_nev_oiii_loiii} compares $\log$([O\,III]/[O\,II]) (top) and $\log([\mathrm{Ne\,V}]/[\mathrm{O\,III}])$ (bottom) against $\log L_{\rm [O\,III]}$, where [Ne~V] denotes the 3427~\AA\ line (\S\ref{sec:intro}). In this analysis, we use the full [Ne\,V] Parent Sample (\S\ref{subsec:nev_parent}), which extends to $z=1.12$, to provide the widest possible dynamic range in AGN luminosity. We use [Ne\,V]/[O\,III] rather than [Ne\,V]/[Ne\,III] because [O\,III] is $\sim13$X stronger than [Ne~III], allowing us to extend the correlation to fainter sources. (Note that in Figure~\ref{fig:ne53_o32_models} A, [Ne\,III] is weaker than [Ne\,V] for 18.4\% of our sample.)
In addition to requiring \NeVp\ S/N $>$ 5, we require S/N $>$ 3 on \OIII\ and \OIIp, yielding a sample of 8,885 galaxies. This is a substantial reduction from the parent sample, largely because [O\,III] is intrinsically weak among many [Ne\,V]-detected galaxies -- consistent with the anti-correlation demonstrated in this section, in which lower-luminosity AGN show progressively stronger [Ne\,V]~relative to [O\,III]. 

Corrections for star formation contamination and dust extinction have not been applied since above $z=0.55$, H$\alpha$ redshifts out of the spectrum, and it is no longer possible to use the BPT to classify star-forming galaxies. The star formation correction primarily impacts [O\,II] at the $\sim20$\% level (see Table~\ref{table:sfg_correction_strength}). 
A dust reddening vector is shown in each panel of Figure~\ref{fig:o32_nev_oiii_loiii} using the median E(B-V) found for the $z < 0.55$ sub-sample. 

To obtain an unbiased view of these line ratio vs. luminosity relations, we supplement the individually [Ne~V]-detected sample (contours in Fig.~\ref{fig:o32_nev_oiii_loiii}) with AGN stacked on a grid of black hole mass and \OIII~luminosity, built without any requirement on [Ne~V]~detection. These stacks are made from the \citet{LawBPT} BPT-AGN sample ($0.147 < z < 0.55$, \S\ref{subsec:nev-ncl_data}) as described in \S\ref{subsec:stacking_methodology}. 
These high S/N stacks avoid the inherent bias toward objects with anomalously strong [Ne~V]/[O~III] when the [Ne\,V] flux nears the detection limit.  To include a more diverse sample, we also utilize the LIER and Composite stacks described in \S\ref{sec:nev-ncl}.  

In Figure~\ref{fig:o32_nev_oiii_loiii}, we see that \ORs~correlates positively with $\log L_{\rm [O\,III]}$, consistent with more luminous AGN having a higher ionization parameter, while $\log$([Ne\,V]/[O\,III])~shows a clear anti-correlation, with the most extreme coronal-to-forbidden line ratios found almost exclusively at the lowest [O\,III] luminosities.  

The stacks agree well with the individual data in the \ORs\ vs.\ $\log L_{\rm [O\,III]}$ plot (Fig.~\ref{fig:o32_nev_oiii_loiii}A), suggesting that sample selection effects have not created any biases in the ratio of these two strong emission lines.  The one exception is at low luminosity ($\log L_{\rm [O\,III]} < 41$) where the AGN stacks lie above the individual data (contours) and the BPT-LIER stacks. The LIERs have stronger low-ionization lines (as their name suggests) and they greatly outnumber low luminosity BPT-AGN in the eBOSS sample (hence the contours and the LIER stacks agree well.)   

In the [Ne\,V]/[O\,III] vs.\ $L_{\rm [O\,III]}$ plot (Fig.~\ref{fig:o32_nev_oiii_loiii}B): the stacks do not agree well with the individual [Ne~V]-detected sample (contours), showing increasingly depressed values of  [Ne\,V]/[O\,III] at progressively lower [O\,III] luminosity.  This suggests that selection effects are important in our individual [Ne~V]-detected sources, especially at low AGN luminosity. Notably, the LIER stacks have slightly higher [Ne\,V]/[O\,III] than the AGN stacks even at fixed [O\,III] luminosity. 

We carry out a weighted linear fit to our AGN, Composite, and LIER stacks with iterative Cook's-distance outlier rejection.  The fits are given below and shown as dashed lines in Figure~\ref{fig:o32_nev_oiii_loiii}.
\begin{equation}
\log([\mathrm{O\,III}]/[\mathrm{O\,II}]) = (0.280 \pm 0.009)~\log L_{\rm [OIII]} - 11.33
\label{eq:o32_loiii_fit}
\end{equation}
\begin{equation} 
\log([\mathrm{Ne\,V}]/[\mathrm{O\,III}]) = (-0.189 \pm 0.012)~\log L_{\rm [OIII]} + 6.73
\label{eq:nev_oiii_loiii_fit}
\end{equation}

\begin{table*}\label{table:nev_highz}
\caption{High Redshift [Ne\,V] $\lambda3427$ Detections 
From the Literature}
\begin{tabular}{||l c c c c c c c l||}
\hline
Identifier & Redshift & log(M$_{*}$) & SFR  & log(sSFR) & log L$_{\rm [O\,III]}$ & [O\,III]/[O\,II] & [Ne\,V]/[O\,III] & Reference \\
  &   & (M$_{\odot}$)  &  (M$_{\odot}$~yr$^{-1}$) & (yr$^{-1}$)  &  (ergs s$^{-1}$) &  &  & \\
\hline\hline
924 & 2.007 & 11.08 & $<47.7$ & $< -9.41$ & 42.004 & 4.226 & 0.1139 & \citet{Valentino2026} \\
329 & 3.064 & 11.11 & $<8.7$ & $< -10.17$ & 42.164 & 1.995 & 0.1274 & \citet{Valentino2026} \\
82 & 3.983 & 10.35 & 21.49 & -9.02 & 42.726 & 6.642 & 0.0718 & \citet{Valentino2026} \\
GN 42437 & 5.587 & 7.9 & 11.6 & -6.83 & 42.609 & 34.15  & 0.0203 & \citet{ChisholmNeV} \\
GS10013609 &  6.931 & 7.7 & 3.9 & -7.11 & 43.278 &  11.46 & 0.0103 & \citet{Scholtz2025} \\ 
JADES-GS-z9-0 & 9.433 & 8.17 & 4.34 & -7.53 & 42.643 & 32.77 &  0.0157 & \citet{Curti2025} \\
\hline \hline
\multicolumn{7}{l}{\footnotesize * The [O\,III] luminosity and the 
line ratios are not corrected for reddening.}
\end{tabular}
\end{table*}

\subsection{Comparison to high redshift}
One important question is whether high-redshift AGN follow similar line ratio--luminosity relations to those shown in Figure~\ref{fig:o32_nev_oiii_loiii}. We have compiled existing literature data where [Ne\,V]~is detected at S/N~$>3$ in Type~II AGN at $z>1$, and plotted it as colored symbols in Figure~\ref{fig:o32_nev_oiii_loiii}. The data include 25 AGN at $z=1.4$--$2.3$ detected in Hubble Space Telescope (HST) grism data from the CLEAR survey \citep{Cleri_MEx}.  These galaxies have stellar masses of $10^{9} - 10^{11.5}$~M$_{\odot}$ and a wide range of SFRs; roughly half of the sample sit near the SFR main-sequence and half above it.  
We also include 6 AGN with James Webb Space Telescope (JWST) NIRSpec data; their properties are reported in Table~\ref{table:nev_highz}. The JWST sample includes 3 AGN at $z=2$--$4$ in massive ($M_* = 10^{10-11}~M_{\odot}$) galaxies that have recently quenched their star formation \citep{Valentino2026}, and 3 low-mass galaxies ($M_* \sim 10^{8}~M_{\odot}$) at $z=5.5$--$9.4$ that harbor extreme starbursts \citep{ChisholmNeV, Scholtz2025, Curti2025}.

The CLEAR comparison sample shows reasonable agreement with the eBOSS data in the [O\,III]/[O\,II] vs. L$_{\rm [O\,III]}$ plot, but sits systematically above our [Ne~V]/[O~III]--$L_{\rm [O\,III]}$ fit. This is likely due to the same selection effect discussed above: individual sources near the survey detection limit will favor AGN with intrinsically strong [Ne\,V] relative to the underlying population at a given luminosity. An additional issue is that the CLEAR spectra were obtained with the HST grism.  At the low spectral resolution of the grism (R=130 - 210), emission lines must have higher EWs to be detected above the continuum. Since [Ne\,V] has lower EW than [O\,III], this bias likely also favors AGN with unusually strong [Ne\,V] relative to [O\,III].

The $z=2-4$ massive post-starburst galaxies \citep{Valentino2026} fall squarely on the relations defined by the eBOSS data. This may be because this comparison sample is very similar in stellar mass and specific SFR (sSFR $\equiv$ SFR/M$_{*}$) to the eBOSS data. (See Table~\ref{table:nev_highz}; for comparision, the eBOSS data has median values of $\log (M_{*}/M_{\odot}) = 11.15$ and log(sSFR/yr$^{-1}$) = -10.3.)

The three extreme high-$z$ starbursts (\citealt{ChisholmNeV, Scholtz2025, Curti2025}) do not lie on the relations defined by the eBOSS data in Figure~\ref{fig:o32_nev_oiii_loiii}. Two of the galaxies show extreme [O\,III]/[O\,II] values that lie more than a dex above our fitted relation.  All three galaxies show strongly depressed [Ne~V]/[O~III] compared to the eBOSS relation. Given the intensity of star formation in these hosts (log(sSFR) = -6.8 -- -7.5), these are likely Composites -- systems where star-forming [O~III]~emission dilutes the AGN [Ne~V]/[O~III]~ratio, as seen for the eBOSS Composites. Indeed, \citet{ChisholmNeV} use photoionization models to show that a $\sim10^{5}$~M$_{\odot}$ black hole contributes roughly 30\% of the H-ionizing photons in GN~42437. Similarly, \cite{Curti2025} finds that JADES-GS-z9-0 falls in the Composite region of several high-$z$ classification diagrams. 

It is worth noting that the JWST starbursts are low-mass galaxies ($M_* \sim 10^8~M_{\odot}$) that likely host black holes with much lower masses than those inferred in most of the eBOSS galaxies. However, models suggest that lower black hole masses enhance [Ne\,V]/[O\,III] relative to [O\,III]/[O\,II] (Fig.~\ref{fig:ne53_o32_models}), so this is unlikely to be the explanation for the depressed position of the high-$z$ starbursts in Figure~\ref{fig:o32_nev_oiii_loiii} B.

In summary, the observed anti-correlation between [Ne\,V]/[O\,III] and L$_{\rm [O\,III]}$ (Fig.~\ref{fig:o32_nev_oiii_loiii} B) reinforces a theme that recurs throughout this paper: extreme ionization states, whether identified via \NeRs\ or [Ne\,V]/[O\,III], are preferentially associated with lower-luminosity, lower-accretion-rate systems rather than the most powerful AGN.
The high-redshift comparison sample offers a first test of whether this relation persists into a regime largely inaccessible to our own survey.  We find that high-$z$ galaxies scatter both above and below the relation, likely due to a combination of selection effects and contributions to the [O\,III] line from star formation.  Thus, caution is advised in using Equation~\ref{eq:nev_oiii_loiii_fit} to predict the expected [Ne\,V] flux in high-$z$ galaxies, especially those with high sSFR. 

One encouraging result of the [Ne\,V]/[O\,III] -- L$_{\rm [O\,III]}$ anti-correlation (Eq.~\ref{eq:nev_oiii_loiii_fit}) is that galaxies that differ in [O\,III] luminosity by 4 orders of magnitude will only differ in [Ne\,V] luminosity by 3.2 orders of magnitude, since less luminous sources will have higher [Ne\,V]/[O\,III]. Thus, despite being a comparatively weak line, [Ne\,V] is a useful AGN diagnostic across a wide range of accretion states.

%------------------------------------------------------------------------------------------------------
%                                                                                                    --
%                                          Section 8                                                 --
%                                                                                                    --
%------------------------------------------------------------------------------------------------------

\section{Comparison to the He~II~BPT Diagnostic} \label{sec:heii}

The He~II~$\lambda$4686/H$\beta$ vs. [N~II]~$\lambda$6584/H$\alpha$ diagram \citep{ShiraziBrinchmann2012} has been proposed as an AGN selection tool well suited to rest-optical spectroscopy at high redshift \citep{Richardson2025}.
He$^{+}$ requires only 54.4~eV to produce, making He~II~a recombination line detectable under substantially softer AGN excitation conditions than [Ne~V]~(97.1~eV).

\begin{figure}[ht]
\plotone{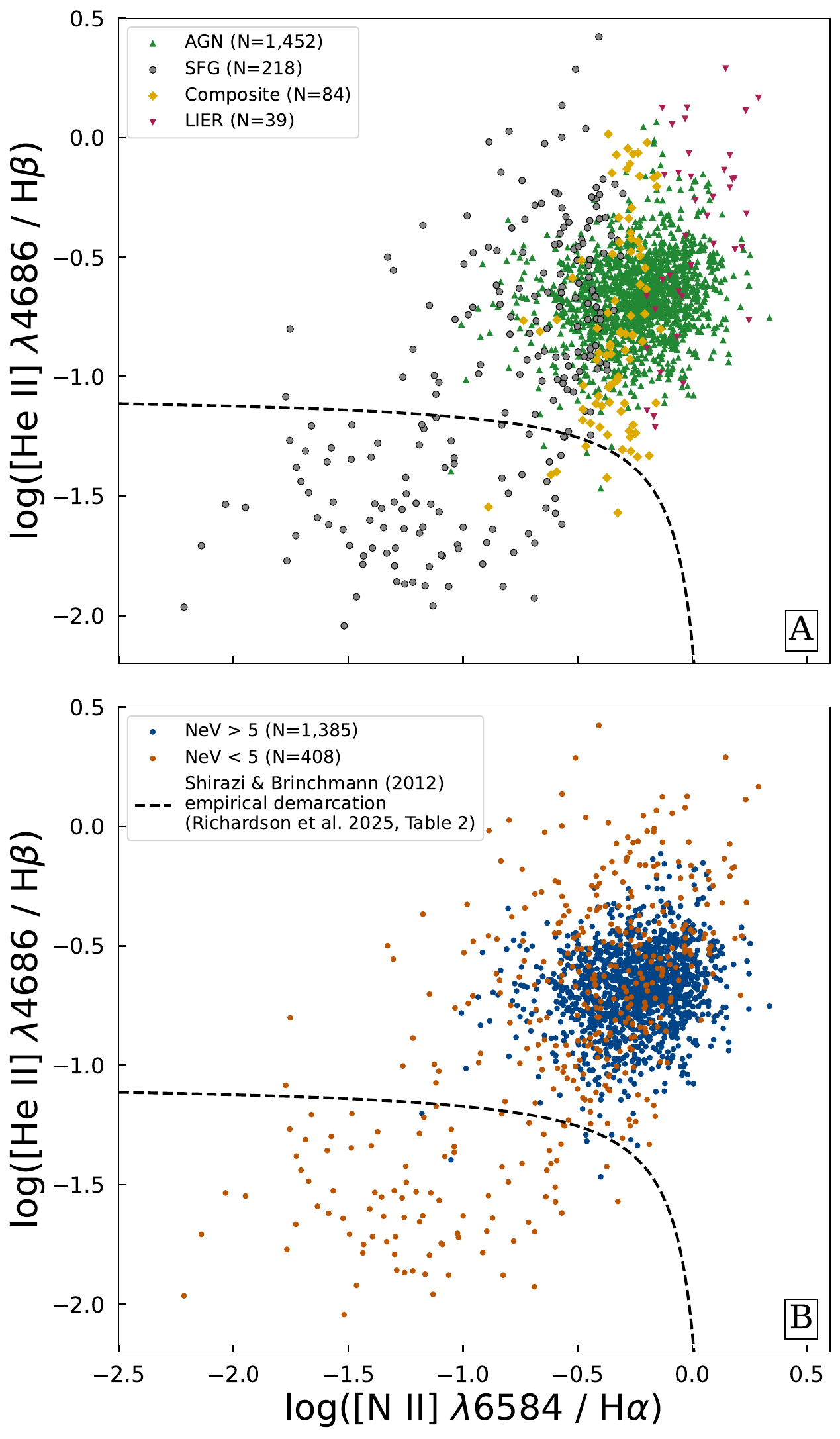}
\caption{The He~II~$\lambda$4686/H$\beta$ AGN diagnostic diagram  for galaxies with $0.147 < z < 0.55$ and He~II, \Ha, \Hb, \NII, and \OIII~all S/N $>$ 3 (N = 1,793).
The dashed line marks the \citet{ShiraziBrinchmann2012} empirical AGN demarcation as quoted in \citet{Richardson2025}.
\textbf{Panel A:} the sample colored by BPT classification: AGN (green triangles, N = 1,452), star-forming (gray circles, N = 218), composite (yellow diamonds, N = 84), and LIER (magenta inverted triangles, N = 39).
\textbf{Panel B:} the same sample, colored by \NeV~detection significance: \NeVp~S/N $>$ 5 (blue, N = 1,385) and \NeVp~S/N $<$ 5 (orange, N = 408).
\label{fig:heii_bpt}}
\end{figure}

Figure~\ref{fig:heii_bpt} shows our BPT sample in the He~II/H$\beta$ diagram, requiring He~II S/N $>$ 3 -- both as a quality cut (matching the S/N $>$ 3 cuts applied to \Ha, \Hb, \NII, and \OIII\ throughout this paper) and as our detection threshold for this single, non-doublet line, since our usual S/N $>$ 5 convention was calibrated for the doublet-fitting boost that [Ne~V] receives and does not directly apply to He~II. The dashed line marks the \citet{ShiraziBrinchmann2012} empirical AGN demarcation line:
\begin{equation} \log(\mathrm{He\,II}/\mathrm{H}\beta) = -1.07 + \frac{1}{(8.92 \log([\mathrm{N\,II}]/\mathrm{H}\alpha) - 0.95)} \end{equation}
as quoted in \citet{Richardson2025}, Table 2. In Panel A, nearly all BPT-classified AGN, Composites, and LIERs fall above the line, as do roughly half of the BPT star-forming galaxies. We stack the 131 BPT-SFG galaxies that fall above the He~II AGN line to test whether this population hosts genuine, individually-undetectable AGN activity; the stack yields no [Ne~V] detection (combined S/N $= 2.28$), well below our S/N $>$ 5 threshold. 
In Panel B, we color code galaxies with and without [Ne~V] detections. All but 5 of the 1,385 [Ne~V]-detected galaxies are correctly classified as AGN by the He~II/H$\beta$ diagnostic, but 316 of the 408 galaxies (77.5\%) without detected [Ne~V] also fall in the AGN region of the diagram. While this seems to suggest that [He~II] is the more sensitive AGN tracer, the opposite is true: among all 3,088 NeV-AGN, 59.6\% lack a He~II detection at S/N $>$ 3, and thus do not appear in the plot.
A He~II-only diagnostic is fundamentally incomplete: the majority of NeV-AGN lack a He~II detection at all. Among He~II-detected sources specifically, the Shirazi \& Brinchmann classification is highly complete (few true AGN are labeled star-forming) but not very pure, given the substantial fraction of BPT star-forming interlopers that fall in its AGN region.

%------------------------------------------------------------------------------------------------------
%                                                                                                    --
%                                          Section 9                                                 --
%                                                                                                    --
%------------------------------------------------------------------------------------------------------

\section{Summary \& Future Prospects} \label{sec:summary}

\NeV~is a promising AGN diagnostic at high redshift, free from the metallicity evolution and ionization-parameter degeneracies that affect traditional optical indicators such as the BPT diagram.
However, the physical conditions that give rise to coronal-line (CL) emission remain poorly characterized, and the small number of confirmed detections in the literature has prevented a rigorous statistical assessment of the purity and completeness of [Ne V]-selected AGN samples.
The eBOSS-DAP catalog \citep{eBOSS-DAP} contains \numNeV~\NeV~detections at S/N $> 5$ across \numNeVgal~galaxies, more than an order of magnitude above all previous literature combined.
We use this sample to characterize the physical conditions that give rise to CL emission and refine [Ne~V] as a diagnostic tool.

Our main findings are as follows:

\begin{enumerate}

    \item Of the 7,433 BPT AGN in our redshift range ($0.147 < z < 0.55$), 3,088 (41.5\%) have \NeVp~S/N $> 5$ (NeV-AGN) and 2,626 (35.3\%) have \NeVp~S/N $< 3$ (NC-AGN), leaving 1,719 (23.1\%) with ambiguous [Ne\,V] status.  Of the [Ne\,V] detections, only 0.9\% are classified as star-forming galaxies in the BPT (Table \ref{table:nev_by_bpt}). Thus, [Ne\,V] provides an extremely pure but very incomplete method of AGN identification.

    \item Of the 3,718 galaxies with \NeVp~S/N $> 5$ across all four BPT classes, 3,088 (83.1\%) are classified as BPT AGN, 598 (16.1\%) are composites or LIERs, and the remaining 32 (0.9\%) are star-forming galaxies (Table~\ref{table:nev_by_bpt}).
    Figure~\ref{fig:bpt_grid} demonstrates that stacking recovers [Ne~V] at S/N $>$ 5 in the large majority of non-star-forming BPT-grid bins (100\% of AGN bins, 90.0\% of LIER bins, and 50.0\% of Composite bins; 87.8\% combined), compared to just 6.7\% of star-forming bins. Equivalent widths in the detected composite and LIER stacks are modest ($\lesssim 5$~\AA). 

    \item Comparing NeV-AGN and NC-AGN directly (Figures \ref{fig:nev-ncl-host-corner}, \ref{fig:nev-ncl-agn-corner}), we find that Eddington ratio and dust content are among the properties that most strongly distinguish the two populations: NeV-AGN accrete at Eddington ratios approximately 2.8 times higher than NC-AGN, while NC-AGN have 55.4\% higher E(B-V) (\S\ref{subsec:host}, \S\ref{subsec:agn}).
    \NeV~therefore preferentially selects rapidly accreting, relatively unobscured AGN and will systematically miss lower-luminosity or dustier systems.
    
    \item Stacking NC-AGN spectra on a grid of black hole mass and \OIII~luminosity (Figure~\ref{fig:mbh-loiii-agn-grid}) and rerunning the eBOSS-DAP on the co-added spectra reveals a formal \NeV~detection in 56 of 83 bins (67.5\%), rising to 83.8\% once bins with fewer than 20 galaxies are excluded, despite every individual parent spectrum falling below our S/N $<$ 3 threshold by construction. This detection rate indicates that the NeV-AGN/NC-AGN dichotomy is primarily a survey sensitivity effect rather than a severe deficit in coronal emission.
    
    \item Comparing [Ne\,V]/[Ne\,III] to [O\,II]/[O\,III] after correcting for star-formation contamination reveals a weak correlation with a large amount of scatter, indicating a diversity of extreme UV SED shapes.  The data are broadly consistent with expectations from AGN photoionization models (Figure~\ref{fig:ne53_o32_models}).

   \item We split AGN in the [Ne\,V]/[Ne\,III]--[O\,II]/[O\,III] diagram into three groups (Fig.~\ref{fig:Ne53-O32-selection}). The ``High53'' group has a lower Eddington ratio, consistent with a radiatively inefficient accretion state analogous to LINERs \citep{Kewley2006}, while the ``Low53'' group consists of rapidly accreting AGN with comparatively small black holes in lower-mass, more highly star-forming hosts than the other two groups. These properties are broadly consistent with a low/hard-to-high/soft accretion continuum, along which hard-continuum, low-\edd\ nuclei produce disproportionately strong high-ionization emission while disc-dominated, high-\edd\ nuclei suppress it.

    \item Roughly 28\% of galaxies in the ``High53'' group in the [Ne\,V]/[Ne\,III]-- [O\,II]/[O\,III] diagram fall outside the BPT AGN region (typically LIERs or Composites), while the fraction is $\lesssim3$\% in the Locus and Low53 groups.  Thus, the $\log$(\NeRs)--\ORs~diagnostic (\S\ref{sec:ner_or}) provides a new method to select samples that contain these elusive lower-luminosity AGN.

    \item The [Ne~V]/[O~III] ratio shows a strong anti-correlation with [O\,III]~luminosity, in contrast to the positive correlation between $\log$([O\,III]/[O\,II]) and [O\,III] luminosity (\S\ref{sec:nev_oiii_loiii}; Figure \ref{fig:o32_nev_oiii_loiii}). 
    The most extreme coronal-to-forbidden line ratios are found almost exclusively among the lowest-luminosity AGN in our sample, reinforcing the picture that extreme ionization states trace lower-accretion-rate systems for the black hole masses that we probe here (M$_{\mathrm{BH}}\sim10^{6-9}$~M$_{\odot}$).

    \item Comparing to a literature sample of six $z=2$--$9$ [Ne~V]-detected AGN from JWST alongside the $z=1.4$--$2.3$ CLEAR sample (Figure~\ref{fig:o32_nev_oiii_loiii}; Table~\ref{table:nev_highz}; \S\ref{sec:nev_oiii_loiii}), we find that some, but not all, high-redshift AGN follow the same relations established at low redshift. Massive, moderately star-forming AGN at $z=2$--$4$ fall squarely on the eBOSS relations, while extreme low-mass starbursts at $z=5.5$--$9.4$ show markedly suppressed [Ne\,V]/[O\,III], likely due to dilution of [O\,III] by star formation.
 
    \item The He~II/H$\beta$ diagram \citep{ShiraziBrinchmann2012} provides a complimentary method to select AGN.  However, use of this diagram alone may miss a large fraction of AGN.  Roughly 60\% of NeV-AGN, 82\% of the full BPT AGN sample, and 95\% of all [Ne~V]-detected galaxies (any BPT class) lack a He~II~detection at S/N $>$ 3 (\S\ref{sec:heii})./

\end{enumerate}

The sample presented in this work provides a powerful low-redshift anchor for future high-redshift studies of the seeding and growth of black holes. Recent works from JWST have failed to detect \NeV\ emission in all but a few $z>5$ galaxies \citep[e.g.,][]{CLeri2023, ChisholmNeV, Scholtz2025, Curti2025}. While heavily driven by selection effects and the shallow spectroscopy in most early JWST surveys, this lack of detections at high redshifts may offer hints toward early black hole growth mechanisms. [Ne\,V] and other coronal line tracers can be a useful tool to disambiguate between heavy direct collapse black hole seeds \citep[e.g.,][]{Bromm2003, Pacucci2023, Pacucci2024a} and potential super-Eddington-accreting light stellar remnant black hole seeds \citep[e.g.,][]{Pacucci2024b,Lambrides2026} by probing the hardness of the ionizing spectrum \citep[c.f.,][]{Richardson2025}. 

The upcoming Nancy Grace Roman Space Telescope \citep{Spergel2015} will be the premier observatory to bridge the gap in cosmic time from our sample to the JWST detections. The Wide Field Instrument (WFI) will provide G150 grism spectroscopy coverage of 1.0-1.93~$\mu$m. G150 traces \NeV\ at $1.92<z<4.63$, covering the peak of cosmic star formation rate density and black hole accretion activity. The High-Latitude Wide Area Survey \citep{Wang2022} Deep tier ($5\sigma$ point source line flux limit $6.2\times10^{-17}$erg s$^{-1}$ cm$^{-2}$) is scheduled to be complete by February 2028. It will provide G150 spectroscopy with an area $\sim1000$x that of comparable grism spectroscopy from HST \citep[e.g.,][]{Cleri_MEx, Simons2023}, which will yield $\sim$1000s of detections of [Ne\,V] at $1.92<z<4.63$.

\section*{Acknowledgments}
OSMA, CAT, and DM gratefully acknowledge support for this work from NSF grant AST-2107725.
OSMA thanks I. McConachie, I. Laseter, M. Maseda, and Z. Lewis for helpful comments.
AD-S gratefully acknowledges support for this work from NSF grant 2107726.
BL gratefully acknowledges support for this work from NSF grant 2107727. 
N.J.C. acknowledges support from JWST-AR-05558, as well as funding from the Eberly Postdoctoral Fellowship from the Eberly College of Science at the Pennsylvania State University. 

This research uses services or data provided by the Astro Data Lab at NSF's NOIRLab. NOIRLab is operated by the Association of Universities for Research in Astronomy (AURA), Inc. under a cooperative agreement with the National Science Foundation.

Funding for the Sloan Digital Sky Survey IV has been provided by the Alfred P. Sloan Foundation, the U.S. Department of Energy Office of Science, and the Participating Institutions. SDSS-IV acknowledges support and resources from the Center for High-Performance Computing at the University of Utah. The SDSS website is available at \href{www.sdss.org}{www.sdss.org}. SDSS-IV is managed by the Astrophysical Research Consortium for the Participating Institutions of the SDSS Collaboration including the Brazilian Participation Group, the Carnegie Institution for Science, Carnegie Mellon University, the Chilean Participation Group, the French Participation Group, Harvard-Smithsonian Center for Astrophysics, Instituto de Astrofísica de Canarias, The Johns Hopkins University, Kavli Institute for the Physics and Mathematics of the Universe (IPMU)/University of Tokyo, Lawrence Berkeley National Laboratory, Leibniz Institut für Astrophysik Potsdam (AIP), Max-Planck-Institut für Astronomie (MPIA Heidelberg), MaxPlanck-Institut für Astrophysik (MPA Garching), Max-PlanckInstitut für Extraterrestrische Physik (MPE), National Astronomical Observatories of China, New Mexico State University, New York University, University of Notre Dame, Observatário Nacional/MCTI, The Ohio State University, Pennsylvania State University, Shanghai Astronomical Observatory, United Kingdom Participation Group, Universidad Nacional Autónoma de México, University of Arizona, University of Colorado Boulder, University of Oxford, University of Portsmouth, University of Utah, University of Virginia, University of Washington, University of Wisconsin, Vanderbilt University, and Yale University.

\section*{Data Availability}\label{sec:data_availability}
The eBOSS-DAP catalog and full documentation are publicly available at \url{https://datalab.noirlab.edu/data/sdss#sdss-iv-eboss-dap-value-added-catalog} \citep{eBOSS-DAP}.
The eBOSS-DAP source code is available at \url{https://github.com/owenmatthewsa/ebossdap}.

\section*{software}
This research made use of
{\tt Astropy}, a community-developed core Python package for Astronomy \citep{astropy:2013,astropy:2018,astropy:2022};
{\tt dust\_extinction} \citep{Gordon2024} for Milky Way foreground extinction correction, using the \citet{GCC09} R(V)-dependent model;
{\tt MaNGA-DAP} \citep{MaNGA-DAP};
{\tt matplotlib} \citep{Hunter:2007};
{\tt numpy} \citep{numpy};
{\tt pandas} \citep{reback2020pandas,mckinney-proc-scipy-2010};
{\tt SciPy} \citep{2020SciPy-NMeth};
{\tt seaborn} \citep{Waskom2021};
and {\tt statsmodels} \citep{seabold2010statsmodels}.

Additionally, computing resources from the University of Wisconsin--Madison Center for High Throughput Computing \citep{CHTC} were used to run the pipeline.

\bibliography{mainbib}{}
\bibliographystyle{aasjournal_notitle}

%-------------------------------------------------------------
\appendix

\section{KS Statistics}\label{apx:ks}

\subsection{Population Comparison (Figures~\ref{fig:nev-ncl-host-corner} and \ref{fig:nev-ncl-agn-corner})}\label{apx:ks:pop}

\begin{table*}[ht]
\centering
\caption{Two-sample KS statistics for the NeV-AGN and NC-AGN populations. $N_{\rm NeV}$ and $N_{\rm NC}$ are given per row, since different properties are subject to different quality cuts (e.g.\ stellar mass/SFR availability, $\sigma_{*}$ quality).}
\label{table:ks_pop}
\begin{tabular}{||l c c c c c c||}
\hline
Property & $N_{\rm NeV}$ & $N_{\rm NC}$ & NeV-AGN Peak & NC-AGN Peak & $D$ & $p$ \\
\hline\hline
Redshift & 3,088 & 2,626 & 0.506 & 0.484 & 0.083 & $7.5 \times 10^{-9}$ \\
$\log(M_{*})$ & 2,971 & 2,545 & 11.121 & 11.206 & 0.130 & $1.4 \times 10^{-20}$ \\
E(B-V) & 3,088 & 2,626 & 0.249 & 0.440 & 0.337 & $3.9 \times 10^{-143}$ \\
$\log(L_{\rm [O\,III]})$ & 3,088 & 2,626 & 42.327 & 41.731 & 0.571 & $\approx 0$ \\
$\sigma_{*}$ (\kms) & 2,633 & 2,345 & 156.134 & 178.424 & 0.072 & $4.8 \times 10^{-6}$ \\
\Ledd & 2,946 & 2,563 & $-1.164$ & $-1.604$ & 0.162 & $6.5 \times 10^{-32}$ \\
\hline
\end{tabular}
\end{table*}

\subsection{Ne-O Three-Group Comparison (Figures~\ref{fig:hll-corner} and \ref{fig:O-Ne-bpt})}\label{apx:ks:neo}

\begin{table*}[ht]
\centering
\caption{Two-sample KS statistics for the three Ne-O groups (N = 316 ``High53'', 1,367 ``Locus'', 315 ``Low53'').}
\label{table:ks_neo}
\begin{tabular}{||l c c c c c||}
\hline
Property & Group 1 Peak & Group 2 Peak & Separation & $D$ & $p$ \\
\hline\hline
\multicolumn{6}{||l||}{\textit{``High53'' vs ``Locus''}} \\
\hline
Redshift & 0.254 & 0.471 & 0.217 & 0.503 & $4.4 \times 10^{-60}$ \\
$\log(M_{*})$ & 11.286 & 11.112 & 0.174~dex & 0.198 & $2.7 \times 10^{-9}$ \\
$\log(\mathrm{SFR})$ & 0.512 & 0.455 & 0.057~dex & 0.152 & $1.2 \times 10^{-5}$ \\
E(B-V) & 0.350 & 0.232 & 0.118~mag & 0.241 & $2.4 \times 10^{-13}$ \\
$\log(L_{\rm [O\,III]})$ & 41.735 & 42.468 & 0.733~dex & 0.562 & $5.8 \times 10^{-76}$ \\
\Ledd & $-1.650$ & $-1.074$ & 0.576~dex & 0.198 & $5.5 \times 10^{-9}$ \\
\hline
\multicolumn{6}{||l||}{\textit{``High53'' vs ``Low53''}} \\
\hline
Redshift & 0.254 & 0.463 & 0.209 & 0.595 & $1.1 \times 10^{-52}$ \\
$\log(M_{*})$ & 11.286 & 10.996 & 0.290~dex & 0.338 & $1.3 \times 10^{-16}$ \\
$\log(\mathrm{SFR})$ & 0.512 & 0.499 & 0.013~dex & 0.186 & $2.6 \times 10^{-5}$ \\
E(B-V) & 0.350 & 0.228 & 0.122~mag & 0.267 & $2.8 \times 10^{-10}$ \\
$\log(L_{\rm [O\,III]})$ & 41.735 & 42.601 & 0.866~dex & 0.655 & $3.4 \times 10^{-64}$ \\
\Ledd & $-1.650$ & $-0.375$ & 1.275~dex & 0.375 & $4.1 \times 10^{-19}$ \\
\hline
\multicolumn{6}{||l||}{\textit{``Locus'' vs ``Low53''}} \\
\hline
Redshift & 0.471 & 0.463 & 0.008 & 0.109 & $4.3 \times 10^{-3}$ \\
$\log(M_{*})$ & 11.112 & 10.996 & 0.116~dex & 0.216 & $5.3 \times 10^{-11}$ \\
$\log(\mathrm{SFR})$ & 0.455 & 0.499 & 0.044~dex & 0.061 & $0.285$ \\
E(B-V) & 0.232 & 0.228 & 0.004~mag & 0.072 & $0.131$ \\
$\log(L_{\rm [O\,III]})$ & 42.468 & 42.601 & 0.133~dex & 0.134 & $1.8 \times 10^{-4}$ \\
\Ledd & $-1.074$ & $-0.375$ & 0.699~dex & 0.222 & $1.4 \times 10^{-10}$ \\
\hline
\end{tabular}
\end{table*}

\end{document}